\documentclass[sigconf]{acmart}
\AtBeginDocument{%
\providecommand\BibTeX{{%
		\normalfont B\kern-0.5em{\scshape i\kern-0.25em b}\kern-0.8em\TeX}}}

\setcopyright{none}
\copyrightyear{2026}
\acmYear{2026}
\setcopyright{none}
\acmConference[WWW '26]{Proceedings of the ACM Web Conference 2026}{April 13--17, 2026}{Dubai, United Arab Emirates}
\acmBooktitle{Proceedings of the ACM Web Conference 2026 (WWW '26), April 13--17, 2026, Dubai, United Arab Emirates}
\acmConference[]{}
\acmBooktitle{-=}
\acmISBN{}
\acmDOI{}

\renewcommand\footnotetextcopyrightpermission[1]{} 

\usepackage[utf8]{inputenc} 
\usepackage[T1]{fontenc}    
\usepackage{booktabs}       
\usepackage{amsmath}
\usepackage{amsfonts}       
\usepackage{nicefrac}       
\usepackage{microtype}      
\usepackage{xcolor}         
\usepackage[normalem]{ulem}
\usepackage{graphicx}

\usepackage{amssymb}
\usepackage{siunitx}
\usepackage{subcaption} 

\usepackage{enumerate}
\usepackage{enumitem}
\usepackage{comment}

\usepackage{bm}

\usepackage{array}
\usepackage{multirow}
\usepackage{threeparttable}
\usepackage{makecell}
\usepackage{tabularx}
\usepackage[procnumbered,ruled,vlined,linesnumbered]{algorithm2e}
\usepackage{tikz}
\usetikzlibrary{arrows.meta,positioning}
\usetikzlibrary{calc}

\newtheorem{problem}{Problem}
\newtheoremstyle{tight}
{0pt}   
{0pt}   
{\itshape} 
{}      
{\bfseries} 
{.}     
{ }     
{}      
\theoremstyle{tight}
\newtheorem{theorem}{Theorem}[section]

\newtheorem{lemma}[theorem]{Lemma}

\newtheorem{definition}[theorem]{Definition}

\def\calG{\mathcal{G}}

\newcommand{\removelatexerror}{\let\@latex@error\@gobble}

\newcommand\LL{\bm{\mathcal{L}}}

\newcommand{\AppxBDRC}[0]{\texttt{AppxBDRC}\xspace}
\newcommand{\PUSH}[0]{\texttt{PUSH}\xspace}
\newcommand{\STW}[0]{\texttt{STW}\xspace}
\newcommand{\SWF}[0]{\texttt{SWF}\xspace}
\newcommand{\ProbeWalk}[0]{\texttt{ProbeWalk}\xspace}

\newcommand\zz{\boldsymbol{\mathit{z}}}

\newcommand\xx{\boldsymbol{\mathit{x}}}

\newcommand\bb{\boldsymbol{\mathit{b}}}

\newcommand\dd{\boldsymbol{\mathit{d}}}

\newcommand\hh{\boldsymbol{\mathit{h}}}
\newcommand\ww{\boldsymbol{\mathit{w}}}

\renewcommand\AA{\boldsymbol{\mathit{A}}}

\newcommand\DD{\boldsymbol{\mathit{D}}}
\newcommand\HH{\boldsymbol{\mathit{H}}}

\newcommand\PP{\boldsymbol{\mathit{P}}}
\newcommand\MM{\boldsymbol{\mathit{M}}}

\newcommand\II{\boldsymbol{\mathit{I}}}

\newcommand\E{\boldsymbol{\mathbb{E}}}
\newcommand\Var{\operatorname{Var}}

\DontPrintSemicolon
\SetKw{KwAnd}{and}
\SetFuncSty{textsc}
\SetKwInOut{Input}{Input\ \ \ \ }
\SetKwInOut{Output}{Output}
\title{ProbeWalk: Fast Estimation of Biharmonic Distance on Graphs via Probe-Driven Random Walks}

\author{Dehong Zheng}
\affiliation{%
	\institution{Fudan University}
	\city{Shanghai}
	\country{China}
}
\email{23307130129@m.fudan.edu.cn}

\author{Zhongzhi Zhang\footnotemark}
\affiliation{
	\institution{Fudan University}
	\city{Shanghai}
	\country{China}
}
\email{zhangzz@fudan.edu.cn}

\begin{document}

\begin{abstract}
The \textit{biharmonic distance} is a fundamental metric on graphs that measures the dissimilarity between two nodes, capturing both local and global structures.
It has found applications across various fields, including network centrality, graph clustering, and machine learning.
These applications typically require efficient evaluation of pairwise biharmonic distances. 
However, existing algorithms remain computationally expensive. 
The state-of-the-art method attains an absolute-error guarantee $\varepsilon_{\mathrm{abs}}$ with time complexity $O(L^{5}/\varepsilon_{\mathrm{abs}}^{2})$, where $L$ denotes the truncation length.
In this work, we improve the complexity to $O(L^{3}/\varepsilon^{2})$ under a relative-error guarantee $\varepsilon$ via probe-driven random walks.
We provide a relative-error guarantee rather than an absolute-error guarantee because biharmonic distances vary by orders of magnitude across node pairs.
Since $L$ is often very large in real-world networks (e.g., $L \ge 10^3$), reducing the $L$-dependence from the fifth to the third power yields substantial gains.
Extensive experiments on real-world networks show that our method delivers $10{\times}$–$1000{\times}$ per-query speedups at matched relative error over strong baselines and scales to graphs with tens of millions of nodes.

\end{abstract}



\maketitle



\section{INTRODUCTION}
In network analysis, graph distance metrics constitute foundational primitives, and numerous variants have been introduced to serve diverse tasks~\cite{yang2022Efficient,lü2011Link,shimada2016Graph,tsitsulin2020Just,jin2019Forest,tran2019Signed,zhang2019Pruning}. An important example is the geodesic distance, which measures distance based on the shortest path between node pairs~\cite{newman2018Networks}. Another notable metric is the resistance distance, arising from the theory of electrical networks, which has proved valuable in diverse domains including circuit theory, chemical graph applications, combinatorial matrix analysis, and spectral graph theory~\cite{doyle1984Random,klein1993Resistance,bapat2014Graphs,yang2013recursion,2007Resistance}.

Among distance measures, the biharmonic distance has emerged as a versatile tool with demonstrated impact in clustering, learning, and network analysis~\cite{yi2018Biharmonica,yi2022Biharmonic,verma2017Hunt,kreuzer2021Rethinking,black2023Understandinga,tyloo2018Robustness,fan2020Spectrala,fitch2016Joint,xu2022Coherence}. The biharmonic distance was originally proposed by Lipman et al.~\cite{lipman2010Biharmonic} for curved-surface geometry, a challenge in computer graphics and geometric processing. Prior work has studied the connection between the biharmonic distance and other measures. For example, \citet{yi2018Biharmonic} connected the biharmonic distance to edge centrality and \citet{wei2021Biharmonic} explored its relationship with the Kirchhoff index.

The biharmonic distance balances neighborhood information with global spectral structure, providing a spectrum-sensitive alternative to geodesic and resistance distances across a range of tasks. Its versatility is exemplified in second-order noisy consensus, where performance metrics naturally scale with the squared distance~\cite{bamieh2012Coherence,yi2022Biharmonic}. Beyond dynamical robustness, the biharmonic distance has found applications in other areas such as graph learning~\cite{kreuzer2021Rethinking,black2023Understandinga}, physics~\cite{tyloo2018Robustness}, and other network-related fields~\cite{fan2020Spectrala,fitch2016Joint,tyloo2019key}.

Despite its importance, computing the biharmonic distance remains a challenging task. Yi et al.~\cite{yi2018Biharmonic} proposed an algorithm for approximating biharmonic distances for all edges, but the algorithm needs to construct a large, dense matrix during preprocessing, which limits its applicability to real-world networks. 
Recently, Liu et al.~\cite{liu2024Fast} developed several efficient algorithms to estimate pairwise biharmonic distance based on random-walk sampling, and one of them, \SWF, is the state-of-the-art algorithm. 
However, the limitation of this algorithm is that on large graphs, it requires a large number of trials to achieve good estimation accuracy, thus its runtime scales as $O(L^{5}/\varepsilon_{\mathrm{abs}}^{2})$. In practice, the truncation length $L$ tends to be large on real-world networks, and thus the $L^{5}$ term in the time complexity results in substantial computational overhead.

These observations motivate an efficient method for evaluating biharmonic distance on graphs with milder dependence on $L$.
Our contributions are summarized as follows:

\begin{itemize}[leftmargin=*]
	\item We introduce a new probe construction, the Paired-Sign Probe, and derive a new formula for the truncated squared biharmonic distance based on this probe.
	
	\item We present \ProbeWalk, an efficient algorithm that achieves time complexity $O(L^{3}/\varepsilon^{2})$ under a relative-error guarantee, reducing the dependence on the truncation length from $L^{5}$ in the state-of-the-art algorithm to $L^{3}$.
	
	\item We conduct extensive experiments on real-world networks and demonstrate that our method delivers \(10{\times}\)–\(1000{\times}\) speedups at matched relative error over state-of-the-art baselines, and that it exhibits strong scalability, remaining practical on graphs with tens of millions of nodes.
\end{itemize}

\section{PRELIMINARIES}

\subsection{Notations}
Throughout this paper, we use bold lowercase (e.g., $\xx$) for vectors and bold uppercase (e.g., $\MM$) for matrices. Entries are indexed by subscripts, e.g., $\xx_i$ or $\MM_{ij}$. We write $\mathbf{1}_i$ for the $i$-th standard basis vector of the appropriate dimension (its $i$-th entry is $1$ and all others are $0$), and $\mathbf{1}$ for the all-ones vector. The transpose of a vector $\xx$ is denoted by $\xx^{\top}$. 

Consider a connected, undirected graph (network) $\calG=(V,E)$ with node set $V$ and edge set $E\subseteq V\times V$. Let $n=|V|$ and $m=|E|$ denote the numbers of nodes and edges, respectively. 
For $i\in V$, let $\mathcal{N}(i)$ be the neighbor set of $i$ and write $\dd_i=|\mathcal{N}(i)|$ for its degree; set $d_{\min}=\min_{i\in V}\dd_i$ and $d_{\max}=\max_{i\in V}\dd_i$. The Laplacian matrix of $\calG$ is a symmetric matrix $\LL=\DD-\AA$, where the adjacency matrix $\AA\in\{0,1\}^{n\times n}$ has entries $\AA_{ij}=1$ if $\{i,j\}\in E$ and $0$ otherwise, and $\DD$ is a diagonal
matrix $\DD=\mathrm{diag}(\dd_1,\ldots,\dd_n)$. Its Moore-Penrose pseudoinverse is $\LL^\dag = \big(\LL +\frac{1}{n}\mathbf{1}\mathbf{1}^\top\,\big)^{-1}-\frac{1}{n}\mathbf{1}\mathbf{1}^\top$. For distinct $u,v\in V$, we define $\bb_{uv}=\mathbf{1}_{u}-\mathbf{1}_{v}$. 
We use $\HH \;=\; \II - \tfrac{1}{n}\,\mathbf{1}\mathbf{1}^\top$ to denote the centering matrix.

We work with the simple random walk on $\calG$. Its transition matrix is $\PP = \DD^{-1}\AA$, so
$\PP_{ij} = \frac{1}{\dd_i}$ if $\{i,j\}\in E$ and $\PP_{ij}=0$ otherwise. For $\ell\ge 0$, let
$(\PP^\ell)_{ij}$ be the probability that a random walk from node $i$ visits node $j$ at the $\ell$-th hop. 
In this paper, following the convention, we assume $\calG$ is not bipartite (which is the case for most real-world networks). 
According to~\cite{motwani1995Randomized}, the random walks over $\mathcal{G}$ are ergodic, i.e., $\lim _{\ell \rightarrow \infty} (\PP^\ell)_{ij}=\bm{\pi}_j = \frac{\dd_{j}}{2 m}$ for any $i, j \in V$, where $\bm{\pi}$ denotes the stationary distribution of a random walk starting from any node.
Writing $\Pi=\mathrm{diag}(\bm{\pi})$, the detailed balance holds: $\Pi\PP=\PP^\top\Pi$~\cite{levin2017Markov}.
Let $\lambda_1\ge \lambda_2\ge\cdots\ge \lambda_n$ be the eigenvalues of $\PP$ with the column eigenvectors
$\ww_1,\ww_2,\ldots,\ww_n$, i.e., $\PP\ww_j=\lambda_j\ww_j$. Let the mixing factor
$\lambda=\max\{|\lambda_2|,|\lambda_n|\}$. Notably, $\lambda_1=1$ and $\ww_1=\mathbf{1}$~\cite{haveliwala2003Second}.

\subsection{Problem Definition}
Since its introduction by Lipman et al.~\cite{lipman2010Biharmonic}, the biharmonic distance on graphs has evolved into a broadly studied spectral distance~\cite{fitch2016Joint,yi2018Biharmonica,yi2018Biharmonic, yi2022Biharmonic,zhang2020Fast,black2023Understandinga,2007Resistance,wei2021Biharmonic}. In intuitive terms, a smaller biharmonic distance indicates two nodes are tightly connected through many routes, whereas a larger biharmonic distance reflects a more indirect connection. Its formal definition is as follows.

\medskip
\begin{definition}\label{def:biharmonic}
	(Pairwise biharmonic distance~\cite{yi2018Biharmonic})
	For a graph $\calG=(V, E)$ with the Laplacian matrix $\LL$, the biharmonic distance $b(s, t)$ between any pair of distinct nodes $s,t\in V$ is defined by
	\begin{align}\label{eq:exact}
		b^{2}(s,t)=\bb_{st}^{\top}\LL^{2\dag}\bb_{st}=\|\LL^{\dag}\bb_{st}\|^2 = \LL_{s,s}^{2\dag} + \LL_{t,t}^{2\dag} - 2\LL_{s,t}^{2\dag}.
	\end{align}
\end{definition}
For notational convenience, we write $\beta(s,t)$ for the squared biharmonic distance $b^{2}(s,t)$ between nodes $s$ and $t$. To avoid redundancy, we will henceforth refer to $\beta(s,t)$ as biharmonic distance instead of the squared biharmonic distance, when it is clear from the context.

\bigskip

\vspace{-0.4em}
\begin{table}[h]
	\centering
	\renewcommand{\arraystretch}{1}
	\fontsize{9}{11}\selectfont
	\caption{Minimum, maximum, and mean $\beta(s,t)$ over 100 uniformly sampled node pairs per network}
	\vspace{-0.3em}
	\begin{tabular}{lrrc}
		\toprule
		Network & Min $\beta(s,t)$ & Max $\beta(s,t)$ & Mean $\beta(s,t)$ \\
		\midrule
		Facebook   &  0.011615 & 7.255681 & 0.840942 \\
		DBLP       &  0.696531 & 3.884803 & 2.129609 \\
		Youtube    &  1.038128 & 4.145025 & 1.881946 \\
		AS-Skitter &  0.030111 & 5.996820 & 1.889589 \\
		Orkut      &  0.000160 & 0.066207 & 0.015191 \\
		LiveJournal&  0.005050 & 5.136179 & 1.117870 \\
		\bottomrule
	\end{tabular}
	\label{tab:max-min-bd}
\end{table}

\vspace{-0.4em}

According to~\cite{black2024Biharmonic}, the biharmonic distance between any distinct nodes $s$ and $t$ satisfies $n^{-2}\leq \beta(s,t)\leq n^3$. This shows that the numerical values of the biharmonic distance can vary drastically across different node pairs. 
To provide a concrete illustration of this dynamic range, we perform a preliminary study on six real-world networks: for each network, we uniformly sample $100$ node pairs and compute their biharmonic distances by running the \PUSH algorithm~\cite{liu2024Fast} for 1000 iterations (achieving a residual below $10^{-6}$). In Table~\ref{tab:max-min-bd}, we show the minimum, maximum, and mean over 100 uniformly sampled node pairs per network.

The theoretical bounds and the empirical summary in Table~\ref{tab:max-min-bd} both show that biharmonic distances span several orders of magnitude across different node pairs. Under such scale variation, an absolute error $\varepsilon_{\mathrm{abs}}$ is not informative: the same $\varepsilon_{\mathrm{abs}}$ can be negligible when $\beta(s,t)\gg \varepsilon_{\mathrm{abs}}$, yet provides no meaningful guarantee when $\beta(s,t)\ll \varepsilon_{\mathrm{abs}}$, because the tolerance dwarfs the true value.

Therefore, in this paper, we focus on approximating pairwise biharmonic distance under a relative-error guarantee, formally defined as follows.
\begin{problem}
	\label{pro:single-pairbd}
	(Pairwise biharmonic distance query) Given a graph $\calG$, a pair of nodes $(s,t)$ with $s \neq t$, and a relative-error guarantee $\varepsilon$, the problem of pairwise biharmonic distance query is to find an estimate $\widehat\beta(s,t)$ such that
	\begin{align}
		\big| \beta(s,t) - \widehat\beta(s,t) \big| \leq \varepsilon \, \beta(s,t).
	\end{align}
\end{problem}

\subsection{Existing Algorithms}

Following its definition, exact computation of biharmonic distance requires computing the Moore-Penrose pseudoinverse $\LL^\dag$ of the Laplacian $\LL$, which can be computed in $O(n^{2.3727})$ time via fast matrix multiplication~\cite{williams2012Multiplying}.
A more practical alternative is to solve Laplacian linear systems $\LL\xx=\bb$ on demand; this primitive has been studied extensively in theoretical computer science, and the fastest nearly-linear algorithms run in $\tilde{O}(m)$ time per solve~\cite{spielman2014Nearly,cohen2014Solving,gao2023Robust}, proving expensive for large-scale graphs. To overcome these difficulties, several approximate approaches have been proposed. 

\paragraph{\AppxBDRC}
Building on Laplacian solvers and random projections, \AppxBDRC approximates biharmonic distance for all edges of a graph $\mathcal{G}$ in time $O(m\log n/\varepsilon_{\mathrm{abs}}^{2})$~\cite{yi2018Biharmonic}.
However, its preprocessing constructs a $(24\log n/\varepsilon_{\mathrm{abs}}^{2})\times n$ embedding matrix for a given $\varepsilon_{\mathrm{abs}}$, which itself requires $\tilde{O}(m/\varepsilon_{\mathrm{abs}}^{2})$ time and a full pass over the graph.
This cost is undesirable when one only needs biharmonic distances for a small set of critical pairs.
Moreover, the method is tailored to edge pairs; ad-hoc queries for arbitrary non-edge node pairs are not supported.

\paragraph{\PUSH\ and \PUSH$+$}
\citet{liu2024Fast} proposed an alternative expression of the biharmonic distance and its truncation bound.

\begin{lemma}[Alternative Expression~\cite{liu2024Fast}]\label{lem:new_formula}
	Let $\hh= \sum_{i=0}^{\infty} \PP^{i} \DD^{-1} \bb_{st}$ where $s,t\in V$ are any two distinct nodes. Then
	\begin{align}
		\beta(s,t) \;=\; \|\hh\|_2^2 \;-\; \frac{1}{n}\bigl(\mathbf{1}^{\top} \hh\bigr)^2.
		\label{eq:beta-basic-expression}
	\end{align}
\end{lemma}

\begin{lemma}[Truncation Bound~\cite{liu2024Fast}]\label{lm:tr}
	Given an integer $L\in[0,\infty]$, for any two distinct nodes $s,t\in V$, let
	$\hh^{(L)} \;=\; \sum_{i=0}^{L} \PP^{i} \DD^{-1} \bb_{st}$ and define
	\begin{align}
		\label{eq:tr-beta-exp-Liu}
		\beta^{(L)}(s,t)
		\;=\; \|\hh^{(L)}\|_{2}^{2}
		\;-\; \frac{1}{n}\bigl(\mathbf{1}^{\top} \hh^{(L)}\bigr)^{2}.
	\end{align}
	For any error $\eta>0$,
	\(
	\bigl|\beta(s,t)-\beta^{(L)}(s,t)\bigr|\le \eta/2
	\)
	whenever the truncation length $L$ satisfies
	\begin{equation}\label{eq:uni_l}
		L \;=\;
		\left\lceil
		\frac{\log\!\bigl( 12n / (\eta(1-\lambda)^2 ) \bigr)}
		{\log(1/\lambda)}
		\right\rceil .
	\end{equation}
\end{lemma}

Starting from Lemma~\ref{lem:new_formula}, the \PUSH algorithm computes the truncated series up to the truncation length $L$, which is chosen according to Eq.~\eqref{eq:uni_l} with the error $\varepsilon_{\mathrm{abs}}$.
Since its time complexity $O(mL)$ scales linearly with the number of edges $m$, the cost quickly becomes prohibitive on large graphs. 
\PUSH$+$ refines the choice of $L$ and improves constants, but the core computation is unchanged and the gain is modest. 

\paragraph{\STW and \SWF}
According to \cite{liu2024Fast}, \STW estimates biharmonic distance via random-walk sampling, and the enhanced \SWF adds variance–aware early stopping with tighter concentration control. 
They work by estimating the \((i,j)\)-step meeting probabilities of random walks, namely the probability that two random walks meet after \(i\) and \(j\) steps, respectively, for \(i,j \in \{0,1,\ldots,L\}\). 
Under an absolute-error guarantee $\varepsilon_{\mathrm{abs}}$ and truncation length $L$, both algorithms run in $O(L^{5}/\varepsilon_{\mathrm{abs}}^{2})$ time.
To the best of our knowledge, \SWF is the state-of-the-art algorithm.
However, the main limitation of these algorithms is their reliance on meeting events between two independent random walks, which ultimately stems from the squared pseudoinverse in the biharmonic distance. Unlike effective resistance $\bb_{st}^\top \LL^{\dagger}\bb_{st}$, the biharmonic distance $\bb_{st}^\top \LL^{2\dagger}\bb_{st}$ contains a square. In walk-based estimators this manifests as products of transition probabilities and, in naive Monte Carlo, is estimated via meeting probabilities between two independent random walks. On large graphs, the \((i,j)\)-step meeting probabilities are vanishingly small; consequently, attaining a given accuracy requires many trials. This explains the \(L^{5}\) dependence in the time complexity. Since \(L\) is typically large (e.g., \(L \ge 10^3\)), this \(L^{5}\) scaling can incur substantial computational overhead.

\section{THE PROBEWALK ALGORITHM}

\begin{figure}[h]
	\centering
	\begin{tikzpicture}
		\node[anchor=south west, inner sep=0] (img)
		{\includegraphics[width=\linewidth]{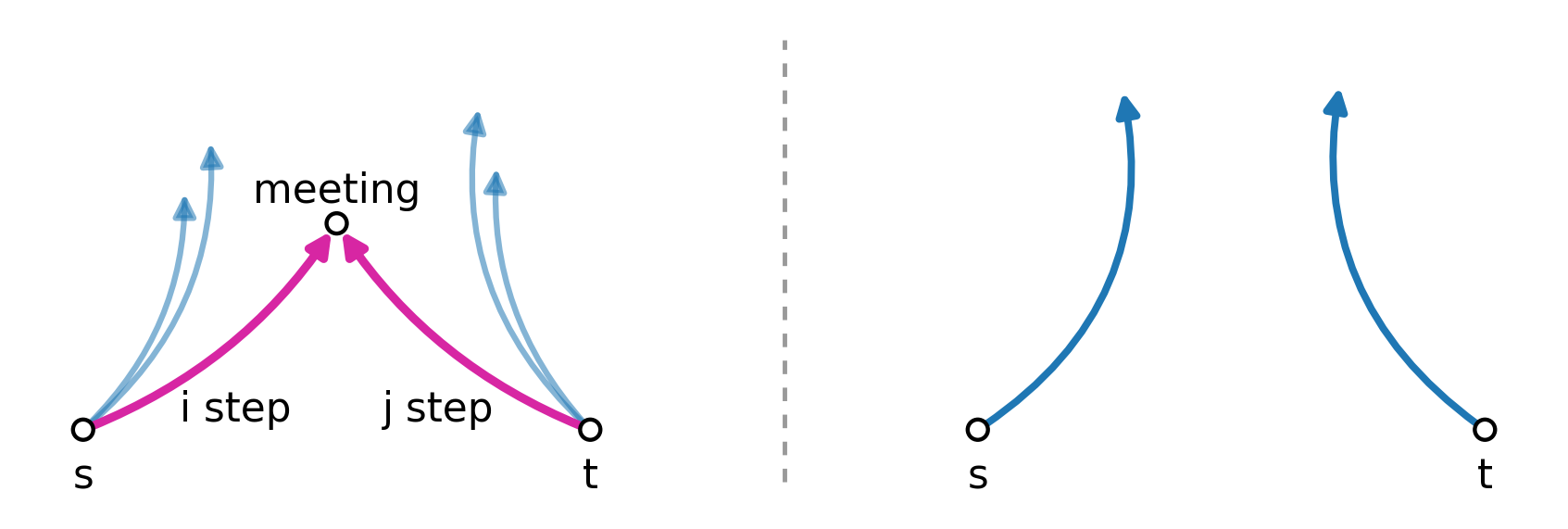}};
		\begin{scope}[x={(img.south east)}, y={(img.north west)}]
			\node[anchor=south, font=\footnotesize\bfseries,
			fill=white, rounded corners=1pt, inner sep=1pt]
			at (0.21, 0.0) {(a)};
			\node[anchor=south, font=\footnotesize\bfseries,
			fill=white, rounded corners=1pt, inner sep=1pt]
			at (0.79, 0.0) {(b)};
		\end{scope}
	\end{tikzpicture}
	\vspace{-2em}
	\caption{\textbf{Meeting-based vs.\ meeting-independent.}
		(a) \STW/\SWF directly estimate the \((i,j)\)-step meeting probabilities;
		(b) \ProbeWalk aggregates local contributions without meetings.}
	\label{fig:intuition-collision}
	\vspace{-0.6em}
\end{figure}

In this section, we present the \ProbeWalk algorithm. We express the truncated squared biharmonic distance as the expectation of a centered quadratic form with respect to a Paired-Sign Probe, yielding $(\zz^\top \hh^{(L)})^2$ with $\E[\zz\zz^\top]=\HH$. This identity enables an estimator that aggregates local contributions along two independent walks from $s$ and $t$, without any meetings. Figure~\ref{fig:intuition-collision} sketches the contrast between prior meeting-based methods and our meeting-independent design.
Then we use a second-order $U$-statistic to remove the squaring bias, and a median-of-means aggregator to provide robust concentration. Consequently, the dependence on the truncation length drops from the fifth power to cubic, yielding an $O(L^{3}/\varepsilon^{2})$ runtime under a relative-error guarantee.

\subsection{New Formula for Biharmonic Distance}
We first introduce the Paired-Sign Probe. Inspired by Rademacher vectors~\cite{achlioptas2003Databasefriendly} and the antithetic-variates idea in Monte Carlo~\cite{robert2004Monte}, the Paired-Sign Probe first zeros out $r$ entries to fix parity, then draws a uniformly random perfect matching on the remaining indices and assigns opposite signs within each pair to enforce pairwise cancellation. Unlike i.i.d. Rademacher probes, for which $\mathbb{E}[zz^\top]=I$, our probe matches a prescribed second moment $\mathbb{E}[zz^\top]=H$.

\medskip

\begin{definition}\label{def:app}
	A Paired-Sign Probe is a random vector $\zz\in\{0,\pm1\}^{n}$ generated as follows.
	Set the parity parameter $r$ by
	\[
	r=\begin{cases}
		1 \,, & \text{if $n$ is odd},\\
		0\text{ or }2\, \text{(each with probability $1/2$)}\, , & \text{if $n$ is even}.
	\end{cases}
	\]
	Draw a uniformly random subset $U\subseteq[n]$ of size $r$ and set $z_u=0$ for each $u\in U$.
	On the remaining index set $[n]\setminus U$ (which has even size), draw a uniformly random perfect matching $M$.
	For each pair $e=\{i,j\}\in M$, sample a base sign $s_e\in\{\pm1\}$ with
	$\Pr(s_e=1)=\Pr(s_e=-1)=\frac{1}{2}$, independently across different pairs, and set
	\[
	\zz_i \;=\; s_e,\qquad \zz_j \;=\; -\,s_e \qquad \text{for each } e=\{i,j\}\in M .
	\]
\end{definition}

\medskip

Based on this definition, the Paired-Sign Probe has a simple second-moment structure.
\begin{lemma}\label{lem:app-second-moment}
	Let $\zz$ be a Paired-Sign Probe.
	Then $\E[\zz_i^2]=1-\frac{1}{n}$, and for $i\neq j$,
	$
	\E[\zz_i \zz_j] = -\frac{1}{n}.
	$
	Equivalently, $\E[\zz\zz^\top]=\HH.$
\end{lemma}

\begin{proof}
	For fixed $r$, a uniformly sampled $U$ gives $\Pr(i\!\in\!U)=r/n$, hence
	\[
	\E[z_i^2]=\Pr(i\notin U)\cdot 1+\Pr(i\in U)\cdot 0 \;=\; 1-\frac{r}{n}.
	\]
	For $i\neq j$, conditioning on $\{i,j\notin U\}$ we obtain
	\begin{align*}
		&\E[z_i z_j \mid i,j\notin U]
		= \Pr(i,j \text{ paired}\mid i,j\notin U)\cdot (-1) \\
		&\quad + \Pr(i,j \text{ in different pairs}\mid i,j\notin U)\cdot 0 
		= -\,\frac{1}{\,n-r-1\,}.
	\end{align*}
	Therefore
	\begin{align*}
		&\E[z_i z_j]
		= \Pr(i,j\notin U)\cdot \E[z_i z_j \mid i,j\notin U] \\
		&= \frac{(n-r)(n-1-r)}{n(n-1)}\;\cdot\;\Bigl(-\frac{1}{\,n-r-1\,}\Bigr) 
		= -\,\frac{(n-r)}{n(n-1)}.
	\end{align*}
	When $n$ is odd, plugging $r=1$ gives $\E[z_i^2]=1-\frac{1}{n}$ and $\E[z_i z_j]=-\frac{1}{n}$.
	When $n$ is even, averaging $r=0$ and $r=2$ with equal weights yields
	\begin{align*}
	&\E[z_i^2] \;=\; \frac{1}{2}\cdot 1 \;+\; \frac{1}{2}\Bigl(1-\frac{2}{n}\Bigr)
	= 1-\frac{1}{n},\\
	&\E[z_i z_j] \;=\; \frac{1}{2}\Bigl(-\frac{1}{n-1}\Bigr) \;+\; \frac{1}{2}\Bigl(-\frac{n-2}{n(n-1)}\Bigr)
	= -\frac{1}{n}.
	\end{align*}
	In both cases we obtain $\E[z_i^2]=1-\frac{1}{n}$ and $\E[z_i z_j]=-\frac{1}{n}$ for $i\ne j$, which is equivalent to $\E[\zz\zz^\top]=\HH$, as claimed.
\end{proof}

\bigskip

Then we reformulate the truncated squared biharmonic distance as the expectation of a centered quadratic form with respect to a Paired-Sign Probe.
\begin{lemma}[New Formula]\label{lm:our-formula}
	Let $\zz$ be a Paired-Sign Probe, $\hh^{(L)} = \sum_{i=0}^{L} \PP^{i} \DD^{-1} \bb_{st}$, and $\phi(\zz)=\zz^\top \hh^{(L)}$. 
	Then
	\begin{equation}\label{eq:beta-app-identity}
		\beta^{(L)}(s,t)
		\;=\;
		\E\big[\phi(\zz)^{2}\big],
	\end{equation}
	where the expectation $\E[\cdot]$ is taken with respect to $\zz$.
\end{lemma}

\begin{proof}
	From Lemma~\ref{lem:new_formula} and Lemma~\ref{lm:tr}, we have
	\begin{align*}
		\beta^{(L)}(s,t)
		= \|\hh^{(L)}\|_{2}^{2} - \frac{1}{n}\bigl(\mathbf{1}^{\top} \hh^{(L)}\bigr)^{2} 
		=\hh^{(L)\top}\HH\hh^{(L)}.
	\end{align*}
	By Lemma~\ref{lem:app-second-moment}, $\E[\zz\zz^\top]=\HH$.
	Substituting this identity into the quadratic form gives
	\begin{align*}
		\beta^{(L)}(s,t)
		= \hh^{(L)\top}\E[\zz\zz^\top]\,\hh^{(L)}
		=\E\big[(\zz^\top \hh^{(L)})^{2}\big]
		=\E\big[\phi(\zz)^{2}\big],
	\end{align*}
	which is exactly Eq.~\eqref{eq:beta-app-identity}, completing the proof.
\end{proof}

\subsection{Random-Walk Based Estimator}\label{subsec:rw-estimator}

We begin by defining \(\widehat{\phi}(\zz)\), an estimator of \(\phi(\zz)=\zz^\top \hh^{(L)}\).
Expanding \(\DD^{-1}\bb_{st}=\mathbf{1}_s/\dd_s-\mathbf{1}_t/\dd_t\) within \(\hh^{(L)}\) yields
\begin{equation}\label{eq:phi-expand-rw}
	\phi(\zz)
	\;=\;
	\sum_{i=0}^{L}\frac{1}{\dd_s}\,\zz^\top \PP^{i}\mathbf{1}_s
	\;-\;
	\sum_{i=0}^{L}\frac{1}{\dd_t}\,\zz^\top \PP^{i}\mathbf{1}_t.
\end{equation}
The detailed balance $\Pi \PP = \PP^\top \Pi$ implies $\Pi \PP^{i}=(\PP^{i})^\top \Pi$ for all $i\ge 0$.
Taking the $(v,u)$-entry of this identity yields
\[
\pi_v\,(P^{i})_{v u} \;=\; \pi_u\,(P^{i})_{u v}\qquad\text{for all }v,u.
\]
Since $\pi_v = d_v/(2m)$, this further implies
\[
d_v\,(P^{i})_{v u} \;=\; d_u\,(P^{i})_{u v}\qquad\text{for all }v,u.
\]
Therefore, for any \(a\in V\) and \(i\ge 0\),
\[
\zz^\top \PP^{i}\mathbf{1}_a
\;=\;
\sum_{v}\zz_v(\PP^{i})_{va}
\;=\;
\dd_a\sum_{v}(\PP^{i})_{av}\frac{\zz_v}{\dd_v}
\;=\;
\dd_a\,\E_{a}\Big[\frac{\zz_{X_i}}{\dd_{X_i}}\Big],
\]
where \(X_i\) denotes the position at step \(i\) of a simple random walk started at \(a\), and \(\E_{a}[\cdot]\) is the expectation with respect to that walk.
Substituting into Eq. \eqref{eq:phi-expand-rw} yields the two-endpoint random-walk identity
\begin{equation}\label{eq:phi-rw-identity}
	\phi(\zz)
	\;=\;
	\sum_{i=0}^{L}\E_{s}\Big[\frac{\zz_{X_i}}{\dd_{X_i}}\Big]
	\;-\;
	\sum_{i=0}^{L}\E_{t}\Big[\frac{\zz_{X_i}}{\dd_{X_i}}\Big].
\end{equation}
Based on this identity, we define an unbiased estimator $\widehat{\phi}(\zz)$ for $\phi(\zz)$, obtained by sampling two independent length-$L$ walks.
Let $X=(X_0=s,X_1,\ldots,X_L)$ and $\widetilde X=(\widetilde X_0=t,\widetilde X_1,\ldots,\widetilde X_L)$ denote independent simple random walks on $\calG$, starting from $s$ and $t$, respectively.
Then we define
\begin{equation}\label{eq:hatphi-def}
	\widehat{\phi}(\zz)
	\;=\;
	\sum_{i=0}^{L}\frac{\zz_{X_i}}{\dd_{X_i}}
	\;-\;
	\sum_{i=0}^{L}\frac{\zz_{\widetilde X_i}}{\dd_{\widetilde X_i}}.
\end{equation}
Conditioning on $\zz$ and using Eq.~\eqref{eq:phi-rw-identity}, each summand has the correct mean; hence $\widehat{\phi}(\zz)$ is an unbiased estimator of $\phi(\zz)$, i.e., \[\E[\widehat{\phi}(\zz)\mid \zz]=\phi(\zz).\]
For notational convenience, hereafter we define the conditional mean and variance of $\widehat{\phi}(\zz)$ given $\zz$ as
\begin{equation}\label{eq:mu-sigma}
\mu \;=\; \E\big[\widehat{\phi}(\zz)\mid \zz\big] \;=\; \phi(\zz),
\quad
\sigma^{2} \;=\; \Var\big(\widehat{\phi}(\zz)\mid \zz\big).
\end{equation}

\medskip
We then construct a second-order U-statistic~\cite{hoeffding1948Class,serfling1980Approximation} from $R$ i.i.d.\ copies of \(\widehat{\phi}(\zz)\) to obtain an unbiased estimator of \(\phi(\zz)^2\).
A straightforward idea is to approximate $\E[\phi(\zz)^2]$ by averaging the squared probes from $R$ independent replicas of Eq. \eqref{eq:hatphi-def}.
However, this plug-in approach leads to bias: conditioning on $\zz$, the identity
\begin{equation}\label{eq:e-phi2}
\E\big[\widehat{\phi}(\zz)^2 \mid \zz\big]
\;=\;
\Var\!\big(\widehat{\phi}(\zz)\mid \zz\big)
\;+\;
\phi(\zz)^2
\end{equation}
shows that the sample average of squared estimates systematically exceeds $\phi(\zz)^2$ by the variance term. 

To remove this variance while keeping the computation simple and streaming-friendly, we use the second-order U-statistic~\cite{hoeffding1948Class,serfling1980Approximation} built from $R$ i.i.d.\ replicas $\widehat{\phi}^{(1)}(\zz),\ldots,\widehat{\phi}^{(R)}(\zz)$ of Eq. \eqref{eq:hatphi-def}, all conditioned on the same $\zz$:
\begin{align}\label{eq:Q-Ustat}
	Q_R(\zz)
	&\;=\;
	\frac{1}{R(R-1)}
	\sum_{p\neq q}
	\widehat{\phi}^{(p)}(\zz)\,\widehat{\phi}^{(q)}(\zz) \\
	&\;=\;
	\frac{\Big(\sum_{p=1}^{R}\widehat{\phi}^{(p)}(\zz)\Big)^{2}
		\;-\;
		\sum_{p=1}^{R}\Big(\widehat{\phi}^{(p)}(\zz)\Big)^{2}}{R(R-1)}.
\end{align}
The cross-product average cancels the diagonal “self-squared’’ terms that carry the variance, leaving only the signal $\phi(\zz)^2$ in expectation.

\smallskip
\begin{lemma}\label{lem:Q-unbiased}
	For any fixed Paired-Sign Probe $\zz$ and any $R\ge 2$,
	\[
	\E\left[\,Q_R(\zz)\,\bigm|\,\zz\,\right]
	\;=\;
	\phi(\zz)^{2}.
	\]
\end{lemma}

\begin{proof}
	We analyze two aggregates of the $R$ replicas separately: the square of their sum and the sum of their squares.
	
	By independence and identical distribution, we decompose the conditional expectation of the squared sum: 
	\begin{align*}
		&\E\left[\!\left(\sum_{p=1}^{R}\widehat{\phi}^{(p)}(\zz)\right)^{2} \Bigm|\!\zz\right]
		= \Var\!\left(\sum_{p=1}^{R}\widehat{\phi}^{(p)}(\zz)\Bigm|\!\zz\right)
		+ \left(\E\left[\sum_{p=1}^{R}\widehat{\phi}^{(p)}(\zz)\Bigm|\!\zz\right]\right)^{2} \\
		&= \sum_{p=1}^{R}\Var\!\left(\widehat{\phi}^{(p)}(\zz)\Bigm|\!\zz\right)
		+ \left(\sum_{p=1}^{R}\E\left[\widehat{\phi}^{(p)}(\zz)\Bigm|\!\zz\right]\right)^{2} 
		= R\,\sigma^{2} + \left(R\,\mu\right)^{2}.
	\end{align*}
	Similarly, we obtain the conditional expectation of the sum of squared replicas:
	\begin{align*}
	&\E\left[\sum_{p=1}^{R}\left(\widehat{\phi}^{(p)}(\zz)\right)^{2}\Bigm|\zz\right]
	\;=\;
	\sum_{p=1}^{R}\E\left[\left(\widehat{\phi}^{(p)}(\zz)\right)^{2}\Bigm|\zz\right] \\
	&\;=\;
	\sum_{p=1}^{R}\left(\Var\!\big(\widehat{\phi}^{(p)}(\zz)\mid \zz\big)\;+\;\left(\E\big[\widehat{\phi}^{(p)}(\zz)\mid \zz\big]\right)^{2} \right) 
	\;=\;
	R\left(\sigma^{2}+\mu^{2}\right).
	\end{align*}
	Subtracting the second equality from the first cancels the variance contribution and leaves $R(R-1)\mu^{2}$.
	Dividing by $R(R-1)$ as in Eq. \eqref{eq:Q-Ustat} gives
	\[
	\E[Q_R(\zz)\mid \zz]
	=\frac{(R\sigma^{2} + (R\mu)^{2})\;-\;R(\sigma^{2}+\mu^{2})}{R(R-1)}
	=\mu^{2}
	=\phi(\zz)^{2},
	\]
	which establishes the unbiasedness claim.
\end{proof}

\medskip
Finally, we aggregate probe estimates via median-of-means for robust averaging over probe randomness.
For any fixed Paired-Sign Probe $\zz$, $Q_R(\zz)$ is unbiased for $\phi(\zz)^2$, but there remains randomness from the draw of $\zz$ itself, since our target quantity $\beta^{(L)}(s,t)$ is an expectation over $\zz$ according to Lemma~\ref{lm:our-formula}. 
Therefore, we draw i.i.d.\ probe vectors $\zz^{(1)},\zz^{(2)},\ldots,\zz^{(K)}$ and form
$\{Q_R(\zz^{(k)})\}_{k=1}^{K}$

A natural choice of aggregation would be the plain average $\frac{1}{K}\sum_{k=1}^{K} Q_R(\zz^{(k)})$. 
Yet in our setting $Q_R(\zz)$ is a second-order U-statistic built from random-walk statistics, and its empirical distribution may contain occasional large values due to rare path realizations. 
The plain average is sensitive to such outliers, and obtaining high-confidence error bounds via classical concentration inequalities may require a large $K$.

To obtain robustness without sacrificing simplicity, we use the median-of-means (MoM) estimator~\cite{nemirovsky1983Problem}. According to~\cite{lugosi2019Mean}, median-of-means delivers sub-Gaussian-style deviation bounds up to absolute constants and is robust to a minority of contaminated or unlucky blocks.

We partition the \(K\) scalars \(\{Q_R(\zz^{(k)})\}_{k=1}^{K}\) into \(G\) disjoint blocks of equal size \(B=K/G\).
Within each block $g\in\{1,2,\ldots,G\}$ compute the block mean
\[
\bar{Y}_g \;=\; \frac{1}{B}\sum_{k\in\mathcal{B}_g} Q_R(\zz^{(k)}).
\]
Our final estimator of $\beta^{(L)}(s,t)$ is given by
\begin{equation}\label{eq:final-estimator}
	\widehat{\beta}^{(L)}(s,t)
	\;=\;
	\mathrm{median}\{\bar Y_1,\bar Y_2,\dots,\bar Y_G\}.
\end{equation}

\subsection{Algorithm Description}\label{subsec:algo}

The \ProbeWalk algorithm is summarized in Algorithm~\ref{alg:rmq}. It first initializes $L,B,G,K,R$ (Line~1). Then, for each probe $\{\zz^{(k)}\}_{k=1}^{K}$, it generates $R$ replicas by simulating paired length-$L$ walks from $s$ and $t$ and updating streaming sums, finally producing a per–probe statistic $Q_R(\zz^{(k)})$ (Lines~2–15). Finally, it partitions $\{Q_R(\zz^{(k)})\}_{k=1}^{K}$ into $G$ blocks, averages within blocks, and returns the median of these $G$ means as $\widehat{\beta}^{(L)}(s,t)$ (Lines~16–20).

\medskip

There is a key operation in our algorithm: generating the probe vectors. If we were to precompute and store all $K$ probe vectors $\{\zz^{(k)}\}_{k=1}^K$, the space and initialization time would both be $\Theta(Kn)$, which would obliterate the advantages of our design and dominate the total running time on large graphs. 
To circumvent this bottleneck, $z^{(k)}_i$ is computed on demand from $(i,k)$ in $O(1)$ time, avoiding explicit materialization of $\zz^{(k)}$.
We now detail the concrete, stateless generator used throughout.

Fix a probe index \(k\) and draw a seedable permutation \(\pi_k:[n]\!\to\![n]\).
We write \(\pi_k^{-1}\) for its inverse permutation.
Both \(\pi_k\) and \(\pi_k^{-1}\) can be evaluated in \(O(1)\) time by instantiating them
with a small-round Feistel pseudorandom permutation~\cite{luby1988How,black2002Ciphers}.

\smallskip
\noindent\textbf{Parity adjustment.}
Let \(r=1\) if \(n\) is odd; if \(n\) is even, set \(r\in\{0,2\}\) by a fair coin deterministically derived from the seed \(k\).
Define the zero set
\[
U \;=\;
\begin{cases}
	\varnothing, & r=0,\\[2pt]
	\{\pi_k^{-1}(n-1)\}, & r=1,\\[2pt]
	\{\pi_k^{-1}(n-2),\,\pi_k^{-1}(n-1)\}, & r=2.
\end{cases}
\]
so that \(|U|=r\) and \(i\in U \iff \pi_k(i)\ge n-r\).
On the \emph{active} indices \([n]\setminus U\), we form pairs exactly as below.

\begin{algorithm}[t]
	\caption{$\ProbeWalk(\calG,s,t, \varepsilon, \delta)$}\label{alg:rmq}
	\DontPrintSemicolon
	\SetAlgoLined
	\KwIn{
		A connected graph $\calG=(V,E)$, an error parameter $\varepsilon$, two nodes $s$ and $t$, a failure probability $\delta$
	}
	\KwOut{Estimated biharmonic distance $\widehat{\beta}^{(L)}(s,t)$}
	\BlankLine
	$L = 
	\left\lceil 
	\frac{\log \left( 
		\frac{96n}{\varepsilon (1-\lambda)^2 (1/d_s+1/d_t)^2}
		\right)}
	{\log(1/\lambda)}
	\right\rceil$;$B=\left\lceil \frac{448}{\varepsilon^{2}} \right\rceil$;
	$G=\left\lceil 8\,\log\!\left(\frac{1}{\delta}\right) \right\rceil$; $K=BG$; $R = \left\lceil
	\frac{128\, (L+1)^2}
	{\,d_{\min}^{2}\!\left(1-\varepsilon/2\right)\left(1/d_s+1/d_t\right)^{2}}
	\right\rceil$;

	\For{$k=1$ \KwTo $K$}{
		\For{$p=1$ \KwTo $R$}{
			$X_0 \leftarrow s$;\quad $\widetilde X_0 \leftarrow t$\;
			$a \leftarrow \zz^{(k)}_{X_0}/\dd_{X_0}$;\quad
			$b \leftarrow \zz^{(k)}_{\widetilde X_0}/\dd_{\widetilde X_0}$\;
			\For{$i=1$ \KwTo $L$}{
				sample $X_i$ uniformly from $\mathcal{N}(X_{i-1})$\;
				sample $\widetilde X_i$ uniformly from $\mathcal{N}(\widetilde X_{i-1})$\;
				$a \leftarrow a + \zz^{(k)}_{X_i}/\dd_{X_i}$;\quad
				$b \leftarrow b + \zz^{(k)}_{\widetilde X_i}/\dd_{\widetilde X_i}$\;
			}
			$U_p \leftarrow a - b$\;
		}
		$S_1 \leftarrow \sum_{p=1}^{R} U_p$;\quad
		$S_2 \leftarrow \sum_{p=1}^{R} U_p^2$\;
		$Q_R(\zz^{(k)}) \leftarrow \dfrac{S_1^2 - S_2}{R(R-1)}$\;
	}
	\BlankLine
	Partition $\{Q_R(\zz^{(k)})\}_{k=1}^{K}$ into $G$ blocks of size $B$;
	
	\For{$g=1$ \KwTo $G$}{
		$\bar Y_g \leftarrow \frac{1}{B}\sum_{k\in\mathcal{B}_g} Q_R(\zz^{(k)})$\;
	}
	$\widehat{\beta}^{(L)}(s,t) \leftarrow \mathrm{median}\{\bar Y_1,\bar Y_2,\dots,\bar Y_G\}$\;
	\Return $\widehat{\beta}^{(L)}(s,t)$
\end{algorithm}

\smallskip
\noindent\textbf{Partner map.}
Define the partner map on active indices by
\[
\rho_k(i)\;=\;\pi_k^{-1}\!\big(\pi_k(i) \oplus 1\big)\qquad\text{for all } i\notin U,
\]
where “\(\oplus 1\)” toggles the least significant bit of \(\pi_k(i)\).
Since toggling the least significant bit maps \(2\lfloor \pi_k(i)/2\rfloor \leftrightarrow 2\lfloor \pi_k(i)/2\rfloor+1\), we have
\[
\{\,i,\rho_k(i)\,\}\;=\;\big\{\,\pi_k^{-1}(2\lfloor \pi_k(i)/2\rfloor),\ \pi_k^{-1}(2\lfloor \pi_k(i)/2\rfloor+1)\,\big\}\subseteq [n]\setminus U .
\]
The collection \(\big\{\{i,\rho_k(i)\}: i\in[n]\setminus U\big\}\) partitions \([n]\setminus U\) into disjoint pairs, which by definition form a perfect matching; with \(k\) as a random seed for \(\pi_k\), this yields a random perfect matching on the active indices.

\smallskip
\noindent\textbf{Base signs.}
Let \(\text{pair}(i)= \lfloor \pi_k(i)/2\rfloor\) be the pair id for \(i\notin U\), so \(\text{pair}(i) \in \{0,1,\dots,(n-r)/2-1\}\). 
By construction,
\[
\operatorname{pair}(\rho_k(i))=\Big\lfloor \frac{\pi_k(i)\oplus 1}{2}\Big\rfloor
=\Big\lfloor \frac{\pi_k(i)}{2}\Big\rfloor
=\operatorname{pair}(i).
\]
We draw a base bit for each pair using a 2-universal hash
\(h_k:\{0,1,\dots,(n-r)/2-1\}\to\{0,1\}\),
\begin{equation*}
	h_k(u) \;=\; \bigl((a_k\,u + b_k)\bmod p\bigr)\bmod 2,
\end{equation*}
where \(p\) is a fixed prime \(> \lfloor n/2\rfloor\).
The parameters \(a_k\in\{1,\dots,p-1\}\) and \(b_k\in\{0,\dots,p-1\}\) are sampled from a seed deterministically derived from \(k\)
(e.g., via a keyed PRF~\cite{luby1988How,black2002Ciphers}).

\smallskip
\noindent\textbf{Entry values.}
Finally, each entry is
\[
z^{(k)}_i 
\;=\;
\begin{cases}
	0, & \text{if } i\in U,\\[3pt]
	(-1)^{\,h_k(pair(i))\,+\,\pi_k(i)}, & \text{if } i\notin U.
\end{cases}
\]
Consequently, each entry \(z^{(k)}_i\) can be computed on the fly in \(O(1)\) time from \((i,k)\), without materializing the full vector \(\zz^{(k)}\).

\medskip

We now analyze the time complexity and the approximation quality of \ProbeWalk. Due to space constraints, we defer the proofs of Lemma~\ref{lem:unconditional-variance} and Theorem~\ref{thm:final} to Appendix~\ref{proof-for-all}. 

\medskip
\begin{lemma}\label{lem:bh-lower-1overd}
	For any pair of distinct nodes $s,t\in V$ with degrees $d_s,d_t$, the squared biharmonic distance
	$
	\beta(s,t)= \bb_{st}^{\top}\LL^{2\dag}\bb_{st}\;
	$
	is bounded below by
	\begin{equation}\label{eq:bh-1overd}
		\beta(s,t)\;\ge\;\frac{1}{8}\Bigl(\frac{1}{d_s}+\frac{1}{d_t}\Bigr)^2.
	\end{equation}
\end{lemma}
\begin{proof}
By definition, $\|\bb_{st}\|_2^2=2$ and $\bb_{st}\perp \mathbf 1$. 
Recall $\beta(s,t)=\bb_{st}^{\top}\LL^{2\dag}\bb_{st}$ and $R_{\mathrm{eff}}(s,t)=\bb_{st}^{\top}\LL^{\dag}\bb_{st}$.
By the Cauchy--Schwarz inequality for positive semidefinite forms~\cite{horn2013Matrix},
\[
\bigl(R_{\mathrm{eff}}(s,t)\bigr)^2
=\bigl(\bb_{st}^{\top}\LL^\dag \bb_{st}\bigr)^2
\le \bigl(\bb_{st}^{\top}\LL^{2\dag}\bb_{st}\bigr)\, \|\bb_{st}\|_2^2
=2\,\beta(s,t),
\]
hence $\beta(s,t)\ge \frac12\bigl(R_{\mathrm{eff}}(s,t)\bigr)^2$.

It has been proved in~\cite{li2023New} that
\[
R_{\mathrm{eff}}(s,t)\;\ge\;\frac{1}{2}\!\left(\frac{1}{d_s}+\frac{1}{d_t}\right).
\]
Therefore,
\[
\beta(s,t)\;\ge\;\frac12\!\left(\frac{1}{2}\!\left(\frac{1}{d_s}+\frac{1}{d_t}\right)\right)^{\!2}
\;=\;\frac{1}{8}\Bigl(\frac{1}{d_s}+\frac{1}{d_t}\Bigr)^{\!2},
\]
which completes the proof.
\end{proof}

\medskip
\begin{lemma}\label{lem:single-projection-variance}
	Let $\sigma^{2} = \Var\big(\widehat{\phi}(\zz)\mid \zz\big)$, and let $L$ denote the truncation length. For any fixed Paired-Sign Probe $\zz$,
	\[
	\sigma^{2}\;\le\;\frac{4(L+1)^{2}}{d_{\min}^{2}}.
	\]
\end{lemma}
\begin{proof}
The estimator $\widehat{\phi}(\zz)$ is a sum of $(L+1)$ differences
\[\Delta_i = \frac{\zz_{X_{i}}}{d_{X_{i}}}-\frac{\zz_{\widetilde X_{i}}}{d_{\widetilde X_{i}}}.\]
Since $\zz_v\in\{0,\pm1\}$ and $d_v\ge d_{\min}$ for all $v$, we have the uniform pointwise bound
\[
|\Delta_i|
\;=\;\Bigl|\frac{\zz_{X_{i}}}{d_{X_{i}}}-\frac{\zz_{\widetilde X_{i}}}{d_{\widetilde X_{i}}}\Bigr|
\;\le\;\frac{1}{d_{X_{i}}}+\frac{1}{d_{\widetilde X_{i}}}
\;\le\;\frac{2}{d_{\min}}
\quad \text{ for every } i.
\]
Hence, 
\[
\bigl|\widehat{\phi}(\zz)\bigr|
=\Bigl|\sum_{i=0}^{L}\Delta_i\Bigr|
\le \sum_{i=0}^{L}|\Delta_i|
\le (L+1)\,\frac{2}{d_{\min}}.
\]
Squaring and taking conditional expectation (given $\zz$) yields
\[
\E\bigl[\widehat{\phi}(\zz)^{2}\mid \zz\bigr]
\;\le\;\Bigl((L+1)\,\frac{2}{d_{\min}}\Bigr)^{2}.
\]
Finally,
$\operatorname{Var}(X)=\E[X^{2}]-(\E[X])^{2}\le \E[X^{2}]$ for any $X$,
so
\[
\sigma^{2} =
\operatorname{Var}\bigl(\widehat{\phi}(\zz)\mid \zz\bigr)
\le \E\bigl[\widehat{\phi}(\zz)^{2}\mid \zz\bigr]
\le \frac{4(L+1)^{2}}{d_{\min}^{2}},
\]
which yields the stated bound.
\end{proof}


\medskip

\begin{lemma}\label{lem:unconditional-variance}
	Let $\theta = \E[\phi(\zz)^2]$. For a second-order U-statistic $Q_R(\zz)$ formed from $R$ i.i.d. replicas of $\widehat{\phi}(\zz)$,
	if $R \ge 4\,\sigma^2/\theta$, then
	\[
	\Var\bigl(Q_R(\zz)\bigr)\;\le\; 7\,\theta^{2}.
	\]
\end{lemma}

We note that the variability of $Q_R(\zz)$ comes from two sources: (i) randomness of the $R$ replicas at a fixed probe $\zz$, and (ii) randomness of the probe $\zz$ itself. This lemma states that once the second-order U-statistic is formed from sufficiently many replicas (specifically, $R\!\ge\!4\sigma^2/\theta$), the overall variability—across both Monte Carlo sampling and the choice of $\zz$—is uniformly bounded by a constant times $\theta^2$.

\medskip

\begin{figure*}[t]
	\centering
	\includegraphics[width=0.98\linewidth]{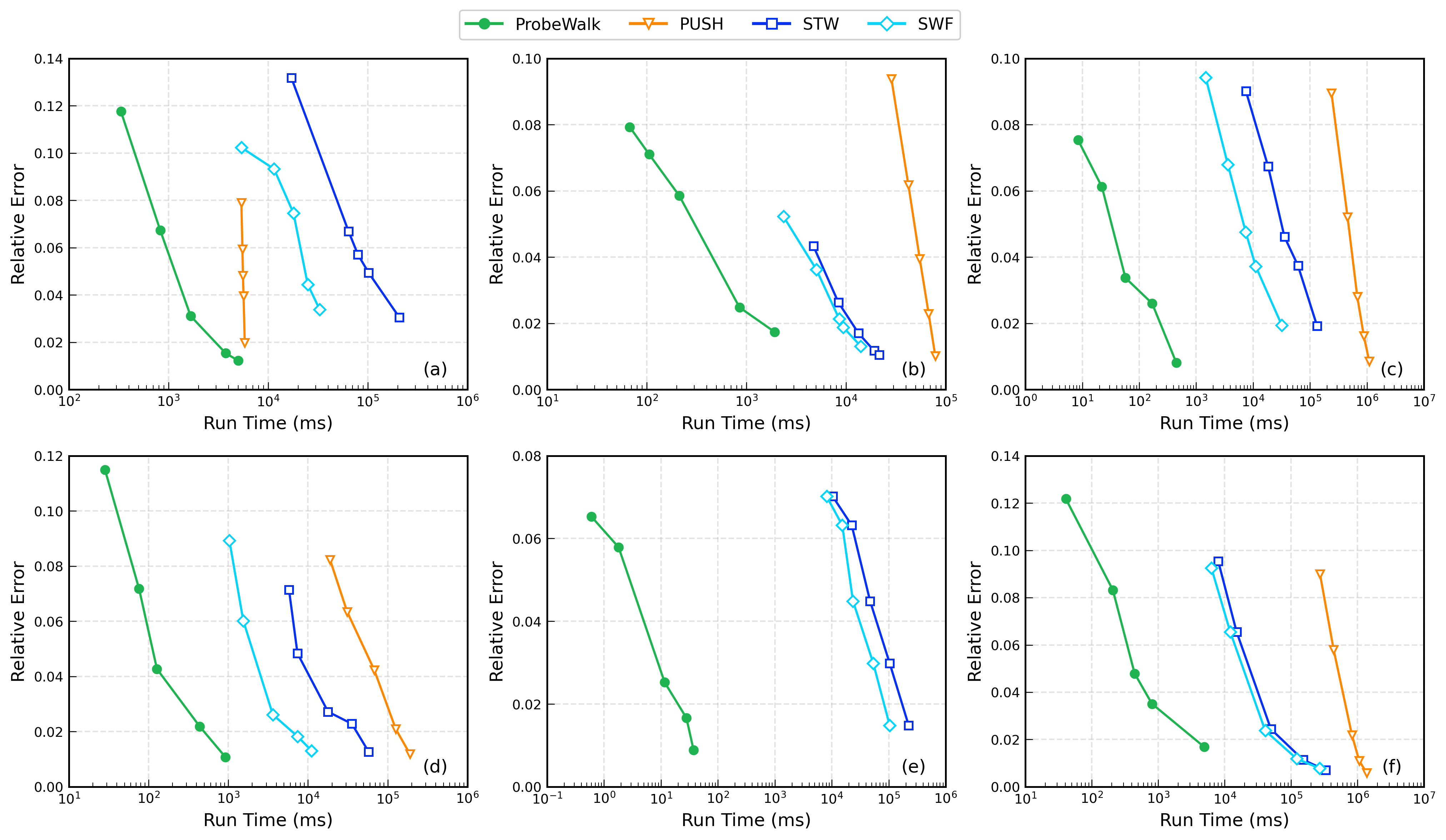}
	\vspace{-0.45cm}
	\caption{Comparison of different algorithms on six real-world networks: (a) Facebook, (b) DBLP, (c) Youtube, (d) AS-Skitter, (e) Orkut, and (f) LiveJournal.}
	\label{fig:running_time}
\end{figure*}

\begin{theorem}\label{thm:final}
	\ProbeWalk returns an estimated biharmonic distance $\widehat{\beta}^{(L)}(s,t)$ such that $\bigl|\widehat\beta^{(L)}(s,t)-\beta(s,t)\bigr|\;\le\;\varepsilon\,\beta(s,t)$  with probability at least $1-\delta$, and the time complexity is bounded by $O\left(\frac{L^3}{\varepsilon^2}\right)$ for $L = 
	\left\lceil 
	\frac{\log \left( 
		\frac{96n}{\varepsilon (1-\lambda)^2 (1/d_s+1/d_t)^2}
		\right)}
	{\log(1/\lambda)}
	\right\rceil$.
\end{theorem}

Lemma~\ref{lem:bh-lower-1overd} provides a degree-only lower bound to parameterize the algorithm with observable features; 
Lemma~\ref{lem:single-projection-variance} and Lemma~\ref{lem:unconditional-variance} control the sampling variance. 
Together they yield the stated relative-error guarantee and runtime bound.

\bigskip

\section{Experiments}\label{sec:exp}

\subsection{Experimental Setup}

\noindent\textbf{Datasets, query sets, and ground truth.}
Our experiments are conducted on nine real-world networks from SOMAR and SNAP~\cite{leskovec2014SNAP,hemphill2018Social}, details of which are provided in Table~\ref{tab:statistics}. For each network, we uniformly sample 100 node pairs as the query set. On the first six networks, we compare \ProbeWalk with existing algorithms to evaluate efficiency and accuracy; ground-truth biharmonic distances for the sampled pairs are computed by running \PUSH\ for 1000 iterations (achieving a residual below $10^{-6}$)~\cite{liu2024Fast}. On the remaining three large-scale networks with tens of millions of nodes, we assess the scalability of our proposed algorithm; other algorithms are infeasible at this scale, so the error control is ensured by the theory of \ProbeWalk\ together with our stabilization rule.

\begin{table}[h]
	\centering
	\renewcommand{\arraystretch}{1}
	\fontsize{9}{11}\selectfont
	\caption{Experimented real-world networks.}
	\vspace{-0.3em}
	\begin{tabular}{lrrc}
		\toprule
		Network & \#nodes ($n$) & \#edges ($m$) & avg($d$) \\
		\midrule
		Facebook   & 4,039     &   88,234     & 43.69 \\
		DBLP       & 317,080   & 1,049,866    & 6.62 \\
		Youtube    & 1,134,890 & 2,987,624    & 5.27 \\
		AS-Skitter & 1,696,414 & 11,095,298   & 13.08 \\
		Orkut      & 3,072,441 & 117,185,083  & 76.28 \\
		LiveJournal& 3,997,962 & 34,681,189   & 17.35 \\
		Bluesky      & 26,692,837 & 1,654,715,214 & 123.98 \\
		Twitter-2010 & 41,652,230 & 1,468,364,884 & 70.51 \\
		Friendster   & 65,608,366 & 1,806,067,135 & 55.06 \\
		\bottomrule
	\end{tabular}
	\label{tab:statistics}
\end{table}

\noindent\textbf{Implementation details.} All experiments are conducted on a Linux machine with an AMD EPYC~7713~(64-core, 2.0\,GHz) processor and 1\,TB of RAM. For fairness and reproducibility, we disable intra-query parallelism and execute each query on a single CPU thread with a uniform memory budget of 32\,GB; no multi-threading or multi-processing is used within a query.
Each experiment loads the graph into memory before timing begins. 
For the first six networks, the spectral quantities $\lambda_2$ and $\lambda_n$ are approximated via ARPACK~\cite{lehoucq1998ARPACK}. 
All algorithms are implemented in Python. 
For randomized algorithms, we set the failure probability $\delta=0.01$, and we exclude a method if it fails to report the result for any query within 24 hours.

\subsection{Query Efficiency on Real-World Networks} \label{subsec:efficiency} 
In this experiment, we aim to assess the efficiency and accuracy of \ProbeWalk compared to \PUSH, \STW and \SWF~\cite{liu2024Fast}.  We exclude \AppxBDRC as a baseline because prior work has shown that \PUSH, \STW and \SWF outperform it by several orders of magnitude.

In Figure~\ref{fig:running_time}, we report the mean wall-clock runtime (ms) and the mean relative error of each method on six real-world networks with error parameter $\varepsilon \in \{0.02, 0.04, 0.06, 0.1, 0.2\}$.
To ensure a fair comparison under a common relative-error guarantee, the truncation length $L$ is set according to Theorem~\ref{thm:final} for each algorithm. The $x$-axis (runtime) is logarithmic, and runtimes are reported in milliseconds. We discuss the main observations below.

Across the six datasets, \ProbeWalk lies on the Pareto front. For a given target accuracy, it attains markedly lower runtime than competing methods; for a fixed time budget, it delivers substantially smaller relative error. On small and medium graphs (a)–(b), \ProbeWalk reaches the $1\%\!-\!2\%$ error region within $10$–$10^3\,$ms, whereas \STW and \SWF require about $10^4\,$ms to achieve similar accuracy. On the larger graphs (c)–(f), the gap widens: the curves for the state-of-the-art algorithm \SWF shift rightward to $10^4$–$10^5\,$ms, while \ProbeWalk remains in the $10^0$–$10^3\,$ms range with comparable or better accuracy. By contrast, \PUSH traces nearly vertical curves: increasing iterations improves accuracy but leaves runtimes in a much higher order of magnitude, so we use it primarily as a high-precision reference. In summary, at matched relative error, \ProbeWalk is $10\times$–$10^3\times$ faster than the state-of-the-art \SWF across all six datasets, establishing a clear efficiency lead.

We further observe that the runtime of \ProbeWalk is not primarily determined by $n$ and $m$. In several cases, larger graphs exhibit shorter per-query times than smaller ones. This can be explained by tighter connectivity of the larger graphs (smaller $\lambda$), which leads to a smaller truncation length $L$.

\vspace{-0.8em}

\subsection{Query Scalability on Large-Scale Networks}
\label{subsec:ultra-10m}

To the best of our knowledge, existing algorithms for biharmonic distance have only been evaluated on graphs with at most a few million nodes, due to their high computational cost. We therefore evaluate \ProbeWalk\ on three large-scale networks—Bluesky, Twitter-2010, and Friendster—each with tens of millions of nodes, under a prescribed accuracy target. On these datasets, existing algorithms fail to return a single pairwise query within 24 hours, whereas \ProbeWalk completes each query within a tractable time budget and yields empirically stable estimates under our stabilization rule.

We fix the target relative error $\varepsilon=0.01$. The truncation length $L$ in Theorem~\ref{thm:final} depends on the spectral quantity $\lambda=\max\{|\lambda_2|,|\lambda_n|\}$; estimating $\lambda$ on these graphs is infeasible within a one-day budget~\cite{lehoucq1998ARPACK}. Instead, we select $L$ via the stabilization rule: starting at $L_0=1$, we double $L$ until the estimate stabilizes, and conservatively accept $L_{\mathrm{stop}}=2L$ if
\[
\big|\widehat\beta^{(2L)}(s,t)-\widehat\beta^{(L)}(s,t)\big|
\;\le\;
\frac{\varepsilon}{2}\cdot \big|\widehat\beta^{(2L)}(s,t)\big|,
\]
otherwise we set $L\!\leftarrow\!2L$ and continue. Empirically, $L_{stop}\le 128$ across all queries.
For completeness, the per-dataset distribution of \(L_{\mathrm{stop}}\) over 100 random queries is deferred to Appendix~\ref{app:lstop-dist}, Figure~\ref{fig:app-lstop-stacked}.

\begin{figure}[t]
	\centering
	\includegraphics[width=.92\linewidth]{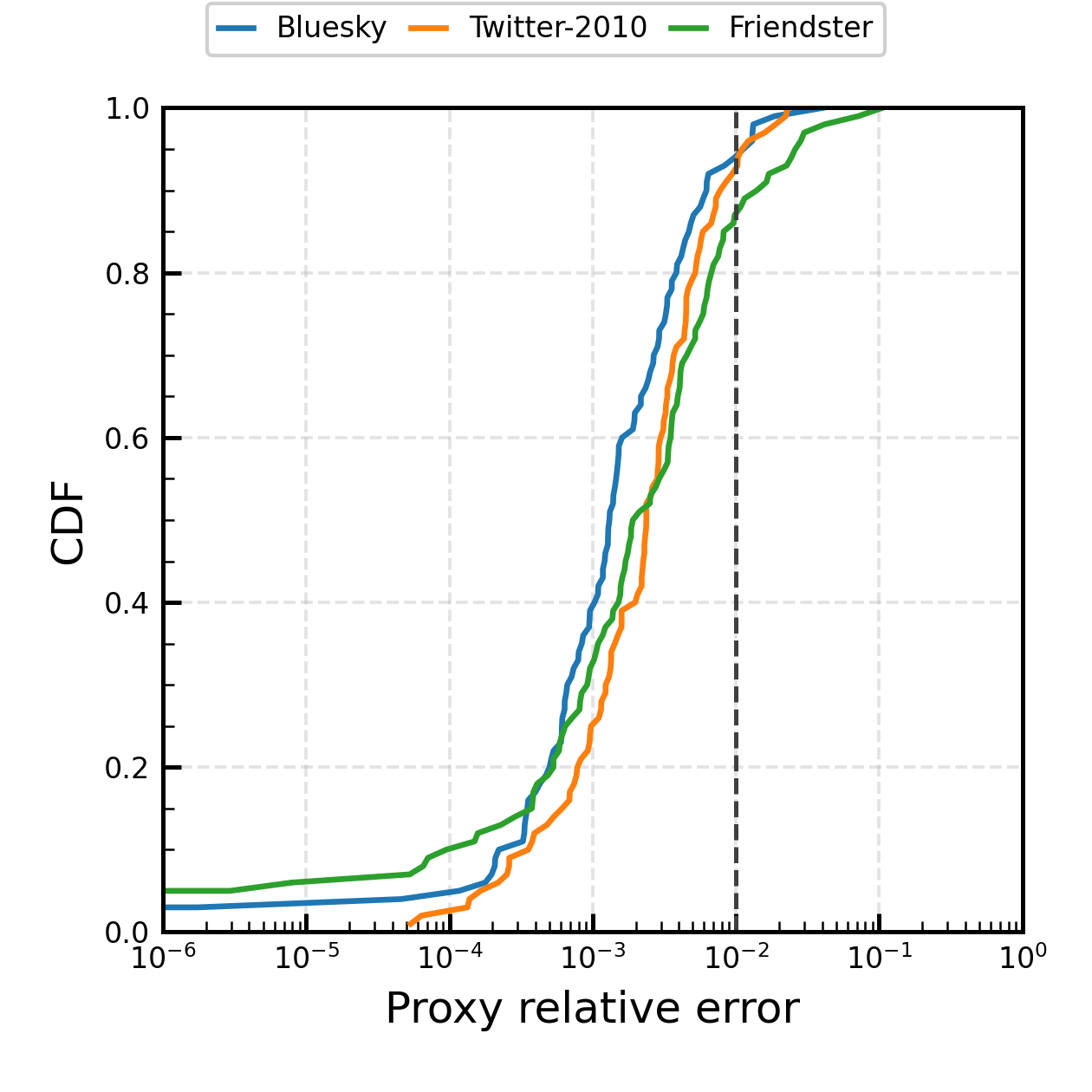} 
	\vspace{-1.5em}
	\caption{
		Cumulative distribution function (CDF) of the proxy relative error over 100 random queries per dataset. The vertical dashed line marks the target $\varepsilon=0.01$}
	\label{fig:ultra-cdf-proxy}
	\vspace{-0.6em}
\end{figure}

To sanity-check that the stabilization rule selects an accurate truncation length on large-scale graphs, we report the proxy relative error
\[
\frac{\big|\widehat\beta^{(L_{\mathrm{stop}})}(s,t)-\widehat\beta^{(L_{\max})}(s,t)\big|}
{\big|\widehat\beta^{(L_{\max})}(s,t)\big|},
\]
where \(L_{\mathrm{stop}}\) is the truncation length chosen by the rule and \(L_{\max}=256\) denotes the maximum truncation length used as the reference.
Figure~\ref{fig:ultra-cdf-proxy} shows the cumulative distribution function (CDF) over the 100 random queries per dataset.
Across Bluesky, Twitter-2010, and Friendster, the vast majority of queries fall below the target \(\varepsilon=0.01\), indicating that the rule typically terminates only after the estimate has stabilized numerically.

\begin{table}[h]
	\centering
	\renewcommand{\arraystretch}{1.0}
	\fontsize{9}{11}\selectfont
	\caption{Per-query runtime (ms) on large-scale graphs: mean, p50, p95, and max over 100 queries.}
	\vspace{-0.3em}
	\begin{tabular}{lrrrr}
		\toprule
		Dataset & mean & p50 & p95 & max \\
		\midrule
		Bluesky      & 34{,}107   & 1{,}398   & 160{,}391   & 1{,}041{,}255 \\
		Twitter-2010 & 54{,}885   & 9{,}943   & 209{,}831   & 1{,}019{,}133 \\
		Friendster   & 301{,}083  & 9{,}404   & 615{,}202   & 6{,}953{,}241 \\
		\bottomrule
	\end{tabular}
	\vspace{-0.6em}
	\label{tab:ultra-10m}
\end{table}

In Table~\ref{tab:ultra-10m}, we report per-query runtimes in milliseconds over 100 queries on large graphs, summarizing the mean, the median (p50), the 95th percentile (p95), and the maximum. The results show that \ProbeWalk completes all 100 queries per dataset, with mean per-query time from 34{,}107~ms to 301{,}083~ms. Runtimes are concentrated near the head: the median ranges from 1{,}398~ms to 9{,}943~ms, and the 95th percentile ranges from 160{,}391~ms to 615{,}202~ms. These results underscore the scalability and practical accuracy of \ProbeWalk on graphs with tens of millions of nodes and billions of edges.

\section{Related Work}

The biharmonic distance originated in geometry processing as a Green's-function-based, spectrum-aware notion of dissimilarity on smooth surfaces~\cite{lipman2010Biharmonic}. Subsequent work transported the concept to graphs by replacing the continuous bi-Laplacian with the squared graph Laplacian, yielding equivalent formulations via the Laplacian pseudoinverse and truncated spectral expansions~\cite{wei2021Biharmonic}. This line of study situates biharmonic distance alongside geodesic and resistance distances, while making explicit links to global indices such as the Kirchhoff index and the biharmonic index~\cite{klein1993Resistance,wei2021Biharmonic}. Recent theory further generalizes biharmonic distance to \(k\)-harmonic variants and analyzes their role in centrality and clustering on graphs~\cite{black2023Understandinga}.

It has proved useful well beyond shape analysis. In dynamical networks, biharmonic distance naturally appears in performance and robustness metrics for second-order noisy consensus and related leader-selection objectives~\cite{bamieh2012Coherence,fitch2016Joint,yi2022Biharmonic,tyloo2018Robustness}. From a learning perspective, the biharmonic distance offers a spectrum-aware metric that emphasizes global structure; together with other spectral distances it underpins positional encodings and spectral clustering, and can be instantiated as a similarity for graph matching—thereby complementing path- and flow-based approaches~\cite{kreuzer2021Rethinking,black2023Understandinga,fan2020Spectrala}.

Algorithmically, exact biharmonic distance computation requires \(\LL^\dag\) or repeated Laplacian solves, which remain heavy on large graphs despite progress on nearly-linear-time SDD solvers~\cite{spielman2014Nearly,cohen2014Solving,gao2023Robust}. 
To improve scalability, \AppxBDRC computes biharmonic distances for all edges via random projections and Laplacian solves~\cite{yi2018Biharmonic}. 
Most recently, a query-oriented line of work formulates single-pair (and nodal) biharmonic distance estimation and proposes push-based and random-walk-based estimators~\cite{liu2024Fast}. However, these methods are primarily analyzed under absolute-error criteria and are computationally expensive on large-scale graphs.

\section{Conclusion}

We introduce a novel probe construction and subsequently propose a new formula for biharmonic distance. Building on this formula, we develop \ProbeWalk, a fast and statistically principled method for pairwise biharmonic distance queries.
The core idea is to rewrite the distance as the expectation of a centered quadratic form under the Paired-Sign Probe. This yields a meeting-independent random-walk estimator: we use two-endpoint, length-$L$ walks without relying on rare meeting events, so the required number of walks is substantially smaller than in meeting-based methods such as \SWF. 
A second-order $U$-statistic removes squaring bias, and a median-of-means aggregator provides robust concentration. 
Analytically, our truncation length dependence improves from the fifth power to cubic order, yielding a significant complexity improvement. 
Extensive experiments on real-world networks corroborate these guarantees, showing consistent accuracy–runtime gains over prior methods by substantial margins.
For future work, it is interesting to examine if our approach can be generalized to the efficient evaluation of the high-order harmonic distances.

\bibliographystyle{ACM-Reference-Format}
\bibliography{refs}

\appendix

\section{EXPERIMENTAL DETAILS}
\subsection{Empirical Degree Distributions}\label{sec:ccdf}

\begin{figure}[h]
	\centering
	\includegraphics[width=\linewidth]{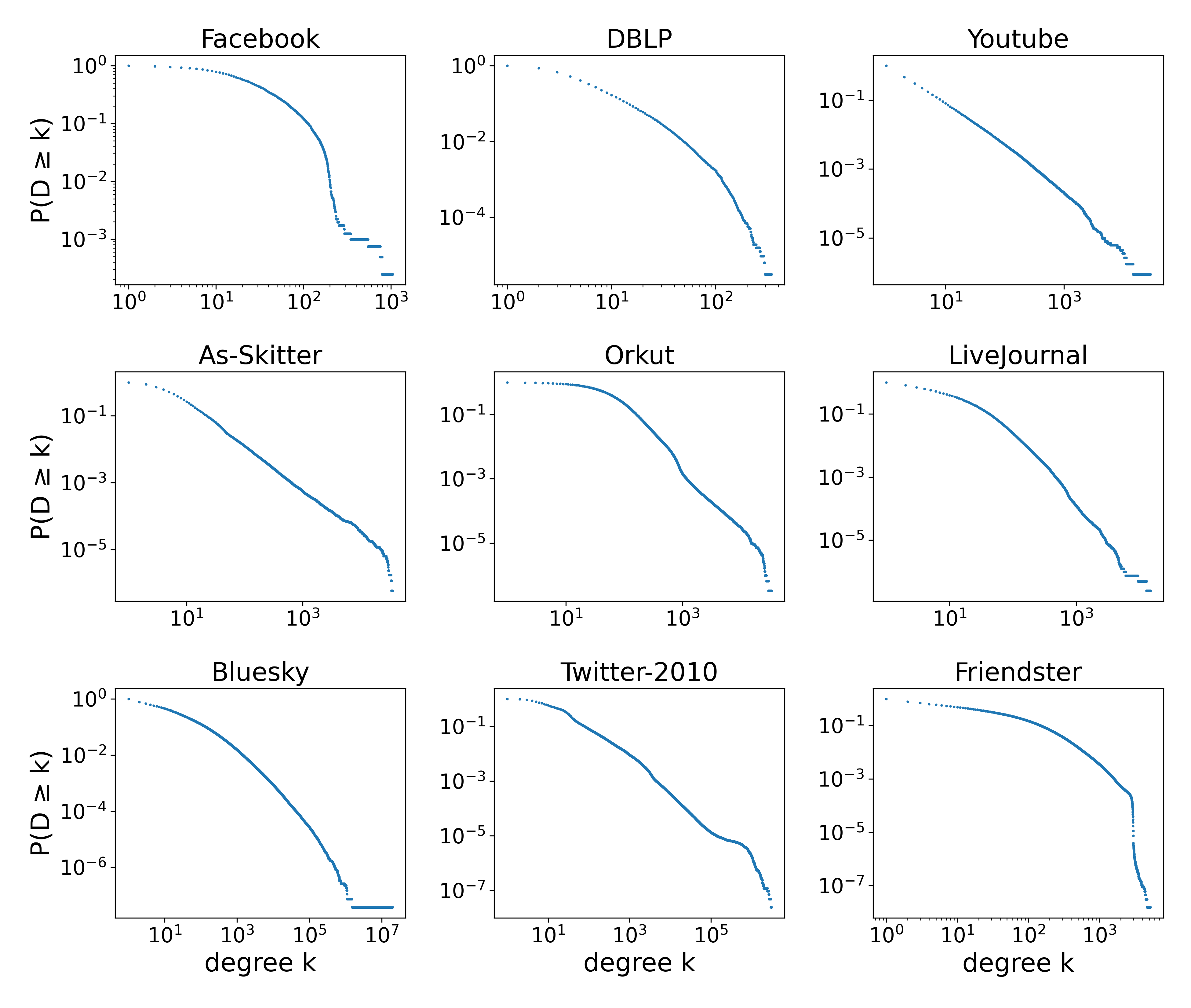}
	\vspace{-1.0em}
	\caption{Complementary cumulative distribution functions of node degree (log–log scale). Each panel plots $P(D\!\ge\!k)$ versus degree $k$ for one network.}
	\label{fig:degree-ccdf-3x3}
	\vspace{-2.0em}
\end{figure}

Figure~\ref{fig:degree-ccdf-3x3} plots the complementary cumulative distribution functions (CCDF) of node degree, defined as
$\Pr\{D\ge k\}$—the fraction of nodes whose degree is at least $k$.
Across all nine networks the distributions are heavy-tailed, but the right tail is extremely light:
once $k$ exceeds a few hundred to a few thousand, $\Pr\{D\ge k\}$ drops by several orders of magnitude and typically falls in the $10^{-3}$–$10^{-7}$ range.
Even in Bluesky and Twitter-2010, where the maximum degree reaches $10^{6}$–$10^{7}$, the CCDF at those levels is about $10^{-6}$–$10^{-7}$, so only a vanishing fraction of nodes are such hubs.
Thus hub–hub queries are very rare in real networks; most pairs in our evaluation contain non–hub nodes.

\subsection{Distribution of selected truncation lengths}
\label{app:lstop-dist}

\begin{figure}[h]
	\centering
	\includegraphics[width=\linewidth]{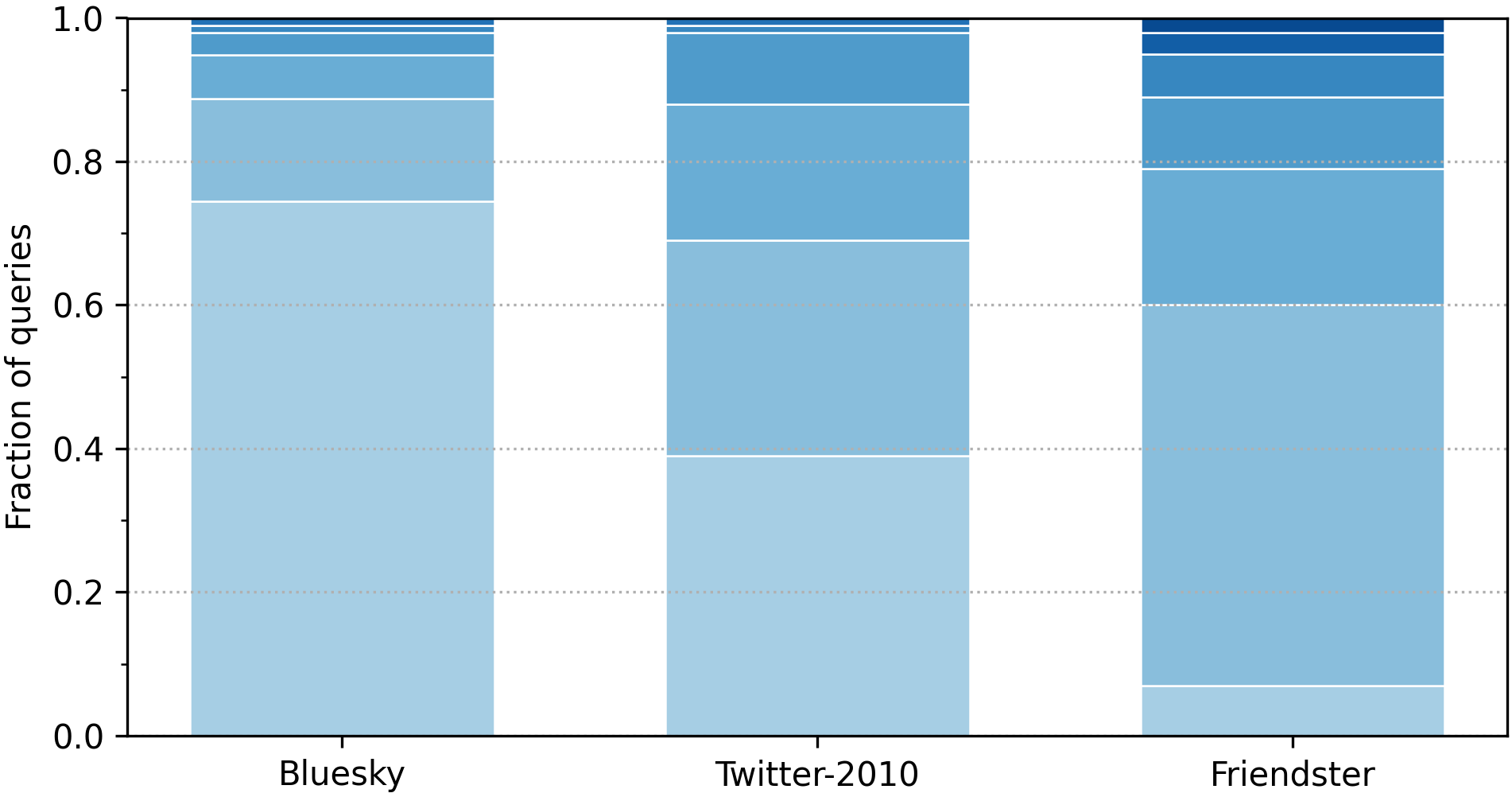}
	\caption{Per-dataset composition of selected truncation lengths across 100 random queries, shown as doublings \(d=\log_{2}(L_{\mathrm{stop}}/L_{0})\). Bars are stacked from bottom to top by increasing \(d\).}
	\label{fig:app-lstop-stacked}
\end{figure}

We visualize the distribution of the selected truncation lengths across the 100 random \((s,t)\) queries for each dataset. Values are expressed in doublings \(d=\log_{2}(L_{\mathrm{stop}}/L_{0})\). Each bar in Figure~\ref{fig:app-lstop-stacked} is stacked from bottom to top by increasing \(d\); higher bands indicate larger truncation lengths.

\section{PROOFS}\label{proof-for-all}

\subsection{Proof of Lemma~\ref{lem:unconditional-variance}}
Our proof follows a sequential structure: we first derive the conditional variance of $Q_R(\zz)$ built from $R$ i.i.d. replicas of $\widehat{\phi}(\zz)$, and then average over random probe vectors $\zz$ to obtain an unconditional upper bound on the variance of $Q_R(\zz)$.

\medskip
\noindent\textbf{Step S1: Conditional variance of the U-statistic.}
Generally, for i.i.d.\ $U_1,\dots,U_R$ and a symmetric kernel $h:\mathbb{R}^2\to\mathbb{R}$,
the variance of the second-order U-statistic
\(
U_R=\binom{R}{2}^{-1}\sum_{p<q} h(U_p,U_q)
\)
admits the standard decomposition ~\cite{hoeffding1948Class}
\[
\Var(U_R)
\;=\; \frac{4}{R}\,\zeta_1 \;+\; \frac{2}{R(R-1)}\,\zeta_2,
\]
where
\[
\zeta_1 \;=\; \Var\bigl(\E[h(U_1,U_2)\mid U_1]\bigr),
\qquad
\zeta_2 \;=\; \Var\bigl(h(U_1,U_2)\bigr) - 2\zeta_1 .
\]

\medskip
In our case, the kernel is $h(u,v)=uv$.
As we have defined before, $U_1=\widehat{\phi}^{(1)}(\zz),U_2=\widehat{\phi}^{(2)}(\zz)$ are i.i.d.\ with mean $\mu$ and variance $\sigma^2$.
First,
\[
\E\bigl[h(U_1,U_2)\mid U_1\bigr]
\;=\; \E[U_1U_2\mid U_1]
\;=\; U_1\,\E[U_2]
\;=\; \mu\,U_1,
\]
hence
\[
\zeta_1 \;=\; \Var(\mu U_1) \;=\; \mu^{2}\sigma^{2}.
\]
Next, since $U_1$ and $U_2$ are independent, and recall that 
$\E[U_1^{2}]=\E[U_2^{2}]=\mu^{2}+\sigma^{2}$ as shown in Eq.~\eqref{eq:e-phi2}, we obtain
\begin{align*}
	&\Var\bigl(h(U_1,U_2)\bigr)
	= \Var(U_1U_2) \\
	&= \E[U_1^{2}]\,\E[U_2^{2}] - \bigl(\E[U_1U_2]\bigr)^{2} 
	= \sigma^{4} + 2\mu^{2}\sigma^{2},
\end{align*}
hence,
\[
\zeta_2
\;=\; \Var(U_1U_2) - 2\zeta_1
\;=\; \sigma^{4} + 2\mu^{2}\sigma^{2} - 2\mu^{2}\sigma^{2}
\;=\; \sigma^{4}.
\]
Substituting $\zeta_1=\mu^2\sigma^2$ and $\zeta_2=\sigma^4$ into the general formula yields
\begin{equation}\label{eq:step1}
\Var\bigl(Q_R(\zz)\mid \zz\bigr)
\;=\; \frac{4}{R}\,\mu^{2}\sigma^{2}
\;+\; \frac{2}{R(R-1)}\,\sigma^{4}.
\end{equation}

\medskip
\noindent\textbf{Step S2: Total variance and the choice of $R$.}
By the law of total variance and Lemma~\ref{lem:Q-unbiased},
\begin{align*}
\Var(Q_R(\zz))
&=\E\bigl[\Var(Q_R(\zz)\mid \zz)\bigr]
+\Var\bigl(\E\left[Q_R(\zz)\bigm|\zz\right]\bigr)  \\
&=\E\bigl[\Var(Q_R(\zz)\mid \zz)\bigr]
+\Var\bigl(\phi(\zz)^2\bigr).
\end{align*}
We bound the two terms separately.

\paragraph{(i) Bounding $\E[\Var(Q_R(\zz)\mid \zz)]$.}
By Eq.~\eqref{eq:step1},
\[
\Var(Q_R(\zz)\mid \zz)\;\le\;\frac{4}{R}\,\phi(\zz)^2\sigma^2+\frac{4}{R^2}\,\sigma^4.
\]
Taking expectations and using $R \ge 4\sigma^2/\theta$ gives
\[
\E\bigl[\Var(Q_R(\zz)\mid \zz)\bigr]
\;\le\;\frac{4\theta \sigma^2}{R}+\frac{4\sigma^4}{R^2}
\;\le\;\theta^2+\theta^2
\;=\;2\theta^2.
\]

\paragraph{(ii) Bounding $\Var(\phi(\zz)^2)$.}
For notational convenience, we define $\alpha=h^{(L)}\in\mathbb{R}^n$, so $\phi(\zz)=\sum_{i=1}^n \alpha_i z_i$.
Under Paired-Sign Probe, the zero set $U$ and the matching $M$ are random; \emph{conditional} on $(U,M)$,
the signs on active pairs are independent Rademacher variables. Write
\[
S(U,M)=\sum_{\{i,j\}\in M}\bigl(\alpha_i-\alpha_j\bigr)^2,\quad |M|=\frac{n-r}{2},
\]
and denote the pairwise differences by $\beta_k$ ($k=1,2,\dots,|M|$). Then
$\phi(\zz)=\sum_{k=1}^{|M|} \beta_k\,\varepsilon_k$ with independent Rademacher $\varepsilon_k$ ($\Pr[\varepsilon_k=1]=\Pr[\varepsilon_k=-1]=\tfrac{1}{2}$). 

By the fact that mixed terms with any odd power $\varepsilon_k$ vanish, we have
\begin{align*}
	&\mathbb{E}\!\left[\phi(\zz)^2\mid U,M\right]
	=\mathbb{E}\!\left[\Bigl(\sum_{k=1}^{|M|}\beta_k\,\varepsilon_k\Bigr)^2\Bigm| U,M\right]\\
	&=\mathbb{E}\!\left[\sum_{k=1}^{|M|}\beta_k^2\,\varepsilon_k^2
	+2\sum_{1\le k<\ell\le |M|}\beta_k\beta_\ell\,\varepsilon_k\varepsilon_\ell \;\Bigm|\; U,M\right]\\
	&=\sum_{k=1}^{|M|}\beta_k^2\,\mathbb{E}[\varepsilon_k^2\mid U,M]
	+2\sum_{1\le k<\ell\le |M|}\beta_k\beta_\ell\,\mathbb{E}[\varepsilon_k\varepsilon_\ell\mid U,M]\\
	&=\sum_{k=1}^{|M|}\beta_k^2\cdot 1
	+2\sum_{1\le k<\ell\le |M|}\beta_i\beta_\ell\cdot 0
	=\sum_{k=1}^{|M|}\beta_k^2 \;=\; S(U,M).
\end{align*}
Similarly, using the identity $\sum_{k=1}^{|M|}\beta_k^2 \;=\; S(U,M)$, we can further write the bound on the fourth moment 
\begin{align*}
	&\E\bigl[\phi(\zz)^4\mid U,M\bigr]
	= \E\left[\Bigl(\sum_{k=1}^{|M|} \beta_k \varepsilon_k\Bigr)^4 \,\middle|\, U,M \right]\\
	&= \sum_{k=1}^{|M|} \beta_k^4 \E[\varepsilon_k^4]
	+ 6\!\sum_{1\le k<\ell\le |M|} \beta_k^2 \beta_\ell^2 \E[\varepsilon_k^2 \varepsilon_\ell^2] 
	= \sum_{k=1}^{|M|} \beta_k^4 + 6\!\sum_{k<\ell} \beta_k^2 \beta_\ell^2 \\
	&= 3\Bigl(\sum_{k=1}^{|M|} \beta_k^2\Bigr)^{\!2} \;-\; 2\sum_{k=1}^{|M|} \beta_k^4
	\;\le\; 3\,S(U,M)^2,
\end{align*}
where we use the identity
$\,(\sum_k \beta_k^2)^2=\sum_k \beta_k^4+2\sum_{k<\ell}\beta_k^2\beta_\ell^2$.

We center $\alpha$ at its mean
$\bar\alpha=\tfrac{1}{n}\sum_{i=1}^n \alpha_i$ and apply the basic inequality
$(u-v)^2 \le 2(u^2+v^2)$, yielding
\begin{align*}
	&S(U,M)
	= \sum_{\{i,j\}\in M}(\alpha_i-\alpha_j)^2 
	= \sum_{\{i,j\}\in M}\bigl((\alpha_i-\bar\alpha)-(\alpha_j-\bar\alpha)\bigr)^2 \\[2pt]
	&\le \sum_{\{i,j\}\in M} 2\Bigl((\alpha_i-\bar\alpha)^2+(\alpha_j-\bar\alpha)^2\Bigr) 
	= 2\sum_{i\in [n]\setminus U}(\alpha_i-\bar\alpha)^2 \\[2pt]
	&\le 2\sum_{i=1}^n(\alpha_i-\bar\alpha)^2 
	= 2\,\alpha^\top H\alpha
	= 2\theta .
\end{align*}
Therefore,
\begin{align*}
	&\E\bigl[\phi(\zz)^4\bigr]
	=\E_{U,M}\!\bigl[\E[\phi(\zz)^4\mid U,M]\bigr]
	\;\le\; 3\,\E_{U,M}\!\bigl[S(U,M)^2\bigr]  \\[-2pt]
	&\;\le\; 3\cdot (2\theta)\cdot \E_{U,M}[S(U,M)]
	=6\theta^2,
\end{align*}
using $S(U,M)\le 2\theta$ and $\E_{U,M}[S(U,M)]=\E[\phi(\zz)^2]=\theta$.
Hence
\[
\Var\bigl(\phi(\zz)^2\bigr)
=\E\bigl[\phi(\zz)^4\bigr]-\E\bigl[\phi(\zz)^2\bigr]^2
\;\le\; 6\theta^2-\theta^2
\;=\; 5\theta^2.
\]

\paragraph{(iii) Combine.}
From (i) and (ii),
\[
\Var(Q_R(\zz))
=\E\bigl[\Var(Q_R(\zz)\mid \zz)\bigr]
+\Var\bigl(\phi(\zz)^2\bigr)
\;\le\;2\theta^2+5\theta^2
\;=\;7\theta^2,
\]
which completes the proof.

\subsection{Proof of Theorem~\ref{thm:final}}

To target a relative-error $\varepsilon$, we apply Lemma~\ref{lm:tr} with $\eta=\varepsilon \beta(s,t)$ and obtain the choice
\[
L_{\mathrm{rel}}  \;=\; 
\left\lceil 
\frac{\log \!\left( 
	\frac{12n}{\varepsilon \beta(s,t)(1-\lambda)^2 }
	\right)}
{\log(1/\lambda)}
\right\rceil.
\]
Since $\beta(s,t)$ is unknown at query time, we plug the lower bound of it in Lemma~\ref{lem:bh-lower-1overd} into $L_{\mathrm{rel}}$, which yields the explicit sufficient choice
\begin{align}\label{L}
	L \;=\; 
	\left\lceil 
	\frac{\log \!\left( 
		\frac{96n}{\varepsilon (1-\lambda)^2 (1/d_s+1/d_t)^2}
		\right)}
	{\log(1/\lambda)}
	\right\rceil.
\end{align}
Since $L \ge L_{\mathrm{rel}}$, our choice of $L$ guarantees
\begin{align}\label{werwefw}
	\bigl|\beta(s,t)-\beta^{(L)}(s,t)\bigr|\;\le\;\frac{\varepsilon}{2}\,\beta(s,t).
\end{align}
In particular, since $\varepsilon<1$,
\begin{align}\label{hetgre}
	\beta^{(L)}(s,t)\;\le\;\beta(s,t)+\frac{\varepsilon}{2}\beta(s,t)\;\le\;2\,\beta(s,t).
\end{align}	

Using the degree–only lower bound (Lemma~\ref{lem:bh-lower-1overd}) together with the truncation guarantee, we have 
\begin{align*}
	&\theta \;=\;\E\bigl[\phi(\zz)^2\bigr] \;=\;\,\beta^{(L)}(s,t)
	\;\ge\;  \Bigl(1-\frac{\varepsilon}{2}\Bigr)\beta(s,t) \\
	&\;\ge\; \frac{1}{8}\Bigl(1-\frac{\varepsilon}{2}\Bigr) \Bigl(\frac{1}{d_s}+\frac{1}{d_t}\Bigr)^{\!2}
\end{align*}
For the single-probe variance, Lemma~\ref{lem:single-projection-variance} yields a bound 
\[
\sigma^2 \;\le\; \frac{4(L+1)^2}{d_{\min}^2}.
\]
For clarity of presentation, we define
\[
\theta_0\;=\;\frac{1}{8}\Bigl(1-\frac{\varepsilon}{2}\Bigr) \Bigl(\frac{1}{d_s}+\frac{1}{d_t}\Bigr)^{\!2}
\]
\[
\sigma_0^{\,2}=\frac{4(L+1)^2}{d_{\min}^2}.
\]
We set
\begin{align}\label{B-value}
	&B \;=\; \Bigl\lceil \dfrac{448}{\varepsilon^2}\Bigr\rceil, \\ \label{G-value}
	&G \;=\; \Bigl\lceil 8\log\bigl(\dfrac{1}{\delta}\bigr)\Bigr\rceil,\\  \label{R-value}
	&R \;=\; \Bigl\lceil \dfrac{4\,\sigma_0^{\,2}}{\theta_0}\Bigr\rceil  = \left\lceil
	\frac{128\, (L+1)^2}
	{\,d_{\min}^{2}\!\left(1-\varepsilon/2\right)\left(1/d_s+1/d_t\right)^{2}}
	\right\rceil,\\ \label{K-value}
	&K \;=\; BG.
\end{align}

By Lemma~\ref{lem:unconditional-variance}, if $R\ge 4\sigma^2/\theta$, then
$\Var\bigl(Q_R(\zz)\bigr)\le 7\theta^2.$
The choice of $R$ given in Eq. \eqref{R-value} implies $R\ge 4\sigma^2/\theta$ because $\sigma^2\le \sigma_0^{\,2}$ and $\theta\ge \theta_0$. Hence, 
\[
\Var(\overline{Y}_g)\;=\;\Var\Bigl(\frac{1}{B}\sum_{k\in\mathcal{B}_g} Q_R(\zz^{(k)})\Bigl)\;\le\;\frac{7B\theta^2}{B^2}=\frac{7\theta^2}{B}.
\]
Chebyshev’s inequality then gives
\[
p \;=\; \Pr\!\left(\,|\bar Y_g-\theta|\ge \frac{\varepsilon\theta}{4}\right)
\;\le\;
\frac{\Var(\bar Y_g)}{(\varepsilon\theta/4)^2}
\;=\; \frac{16\cdot 7\theta^2}{B\varepsilon^2\theta^2}
\;=\; \frac{1}{4},
\]
where the last equality uses $B$ in Eq. \eqref{B-value}.

Let $U=\sum_{g=1}^{G}\mathbf{1}\{|\bar Y_g-\theta|\ge \varepsilon\theta/4\}$, so that $U\sim\mathrm{Bin}(G,p)$ with $p\le 1/4$.
By Hoeffding's inequality (one-sided) for Bernoulli sums~\cite{hoeffding1963Probability},
\begin{align*}
	&\Pr(U\ge G/2)
	= \Pr\!\left(\frac{U}{G}-p \ge \frac{1}{2}-p\right) \nonumber\\
	&\le e^{\left(-2G\big(\frac{1}{2}-p\big)^2\right)}
	\ \le\ e^{-G/8}. \label{eq:Uhalf}
\end{align*}
If $U<G/2$, then strictly more than half of $\{\bar Y_g\}$ lie in $[\theta-\varepsilon\theta/4,\theta+\varepsilon\theta/4]$, forcing the median to lie in this interval; hence the event
$
\Bigl|\mathrm{median}\{\bar Y_1,\dots,\bar Y_G\}-\theta\Bigr| \;\ge\; \frac{\varepsilon\theta}{4}
$
implies $\{U\ge G/2\}$, it follows that
\begin{equation}
	\Pr\!\left(\Bigl|\mathrm{median}\{\bar Y_1,\dots,\bar Y_G\}-\theta\Bigr| \;\ge\; \frac{\varepsilon\theta}{4}\right)
	\ \le\ \Pr(U\ge G/2)
	\ \le\ e^{-G/8}. \label{eq:median_bound}
\end{equation}
Therefore, the choice of $G$ in Eq. \eqref{G-value} guarantees
\begin{equation}\label{afadfaw}
	\Pr\!\left(\Bigl|\mathrm{median}\{\bar Y_1,\dots,\bar Y_G\}-\theta\Bigr|\ \ge\ \frac{\varepsilon\theta}{4}\right)\ \le\ \delta.
\end{equation}

Because $\theta=\beta^{(L)}(s,t)$ and $\mathrm{median}\{\bar Y_1,\dots,\bar Y_G\}
=\widehat{\beta}^{(L)}(s,t)$, we have
\[
\bigl|\widehat\beta^{(L)}(s,t)-\beta^{(L)}(s,t)\bigr|
\;\le\; \frac{\varepsilon}{4}\,\beta^{(L)}(s,t),
\quad\text{with probability at least }1-\delta.
\]
Combining this with the truncation bound Eq. \eqref{werwefw} and using $\beta^{(L)}(s,t)\le 2\,\beta(s,t)$ gives
\[
\bigl|\widehat\beta^{(L)}(s,t)-\beta(s,t)\bigr|
\;\le\; \frac{\varepsilon}{4}\cdot 2\,\beta(s,t) + \frac{\varepsilon}{2}\,\beta(s,t)
\;\le\; \varepsilon\,\beta(s,t),
\]
with probability at least $1-\delta$, as claimed.

With $K=\Theta\!\bigl(\varepsilon^{-2}\log(1/\delta)\bigr)$ and $R=\Theta\left(L^2/((1/d_s+1/d_t)^2 d_{\min}^2)\right)$, the total work satisfies
\[
\Theta\bigl(KRL\bigr)
\;=\;
\Theta\Bigl(\frac{\log(1/\delta)}{\varepsilon^2}\Bigr)\cdot 
\Theta\Bigl(\frac{L^2}{(1/d_s+1/d_t)^2 d_{\min}^2}\Bigr)\cdot \Theta(L)
\;=\; O\Bigl(\frac{L^3}{\varepsilon^2}\Bigr),
\]
We suppress in the big-$O$ notation an instance-dependent factor 
$\Theta\bigl(\log(1/\delta)/\bigl((1/d_s+1/d_t)^2 d_{\min}^{\,2}\bigr)\bigr)$. 
Using $(1/d_s+1/d_t)\!\ge 2/d_{\max}$, this factor is at most 
$O\bigl(\log(1/\delta)\,(d_{\max}/d_{\min})^{2}\bigr)$ in the worst case.
But in practice, we fix $\delta\!=\!0.01$, and in real-world networks only a small fraction of nodes are high-degree hubs (see Appendix~\ref{sec:ccdf}, Figure~\ref{fig:degree-ccdf-3x3}), so the event that either endpoint is a hub has low probability; correspondingly, this instance factor is typically much smaller than its worst-case bound.

\end{document}